\documentclass[useAMS,usenatbib,usegraphicx]{mn2e}
\usepackage{amsmath,amssymb}
\usepackage{epsfig,subfigure,color}
\usepackage{multirow}
\usepackage[T1]{fontenc}
\usepackage{ae,aecompl}

\newcommand{\vecg}{\ensuremath{{\bmath{g}}}}
\newcommand{\vecU}{\ensuremath{{\bmath{U}}}}

\newcommand{\vecnab}{\ensuremath{{\bmath{\nabla}}}}

\def\gb {\vecg}
\def\Ub {\vecU}

\newcommand{\nyx}{\textsc{Nyx}}

\newcommand{\sub}[2]{\ensuremath{#1_{\text{#2}}}}
\newcommand{\unit}[2]{\ensuremath{#1\,{\text{#2}}}}

\title[Hot and Turbulent Gas in Clusters]{Hot and Turbulent Gas in Clusters}
\author[W. Schmidt et al.]{\parbox{18cm}{
        W.~Schmidt$^{1,2}$\thanks{E-mail:wolfram.schmidt@uni-hamburg.de}, 
        J.~F.~Engels$^{2}$,
        J.~C.~Niemeyer$^{2}$,
        A.~S.~Almgren$^{3}$}\vspace{0.3cm}\\
$^{1}$Hamburger Sternwarte, Universit\"at Hamburg, Gojenbergsweg 112, D-21029 
Hamburg, Germany\\
$^{2}$Institut f\"ur Astrophysik, Universit\"at G\"ottingen, Friedrich-Hund-Platz 1, D-37077 G\"ottingen, Germany\\
$^{3}$Center for Computational Sciences and Engineering, Lawrence Berkeley National Laboratory, Berkeley, CA 94720, USA}

\begin{document}

\date{Revised draft version}

\pagerange{\pageref{firstpage}--\pageref{lastpage}} \pubyear{2013}

\maketitle

\label{firstpage}

\begin{abstract}
The gas in galaxy clusters is heated by shock compression through accretion (outer shocks) and mergers (inner shocks). These processes additionally produce turbulence. 
To analyse the relation between the thermal and turbulent energies of the gas under the influence of non-adiabatic processes, we performed numerical simulations of cosmic structure formation in a box of 152 Mpc comoving size with radiative cooling, UV background, and a subgrid scale model for numerically unresolved turbulence. By smoothing the gas velocities with an adaptive Kalman filter, we are able to estimate bulk flows toward cluster cores. This enables us to infer the velocity dispersion associated with the turbulent fluctuation relative to the bulk flow. For halos with masses above $10^{13}\,M_\odot$, we find that the turbulent velocity dispersions averaged over the warm-hot intergalactic medium (WHIM) and the intracluster medium (ICM) are approximately given by powers of the mean gas temperatures with exponents around 0.5, corresponding to a roughly linear relation between turbulent and thermal energies and transonic Mach numbers. However, turbulence is only weakly correlated with the halo mass. Since the power-law relation is stiffer for the WHIM, the turbulent Mach number tends to increase with the mean temperature of the WHIM. This can be attributed to enhanced turbulence production relative to dissipation in particularly hot and turbulent clusters.
\end{abstract}


\section{Introduction}

In astrophysics, adiabatic and isothermal turbulence are two important cases, corresponding to the limits of negligible cooling on the one hand and extremely efficient cooling through radiation on the other hand. In the context of the interstellar medium, the properties of isothermal compressible turbulence have been thoroughly studied. In particular, many simulations of forced turbulence in  periodic boxes were performed to infer scaling relations and other statistics \citep[e.~g.][]{KritNor07,SchmFeder09a,FederRom10,KonstShet15}. In the isothermal case, the dissipated energy is instantaneously radiated away. This results in a fixed ratio between the mean turbulent and thermal energies, in which equilibrium between energy injection and dissipation is maintained. In particular, this implies a nearly constant root mean square (RMS) Mach number of the turbulent flow. Forced isothermal turbulence was revisited for an expanding space with constant Hubble parameter $H$ by \citet{SchmAlm14}. As long as the dynamical time scale of turbulence is short compared to the Hubble time $1/H$, however, the RMS Mach number remains steady. 

In contrast to isothermal turbulence, stirring and dissipation in adiabatic gas steadily raise the thermal energy, which in turn causes a rapidly declining Mach number  \citep[see][]{SchmHille06,SchmAlm14}.
The assumption of adiabaticity was often used to approximate the dynamics of the baryonic component in simulations of cosmic structure formation. For example, mergers of dark-matter halos excite large-scale motions through tidal forces, which are subsequently dissipated. Moreover, gas that is accreted from filaments and from the dilute medium in the void experiences strong shock compression when it hits the denser medium inside a galaxy cluster (the basic physics of shock-generated turbulence is briefly reviewed in Section~\ref{sc:shock-gener-turb}). As a result, accretion and mergers heat the baryonic gas to temperatures up to a few keV. These processes have been studied in a variety of cosmological simulations \citep[e.~g.][]{Rick98,NorBry99,DolVaz05,MaierIap09,VazzaBrun09b,IapSchm11,PaulIap11,VazzaBrun11,Miniati2014}.
However, the UV background radiation produced by galaxies acts as a temperature floor, thus limiting the shock strength compared to the adiabatic case \citep{QuinnLidz09,VazzaBrun09a}. At higher densities, radiative cooling reduces the temperature of shock-heated gas, particularly in the warm-hot intergalactic medium (WHIM), but also in the intracluster medium (ICM) \citep{KatzWein96,FaltKarvt05,KangRyu07,NagKravt07,LauKravt09,BorKravt11,SmithHall11,IapViel13,LukStark15,HahnMar15}.
Since cooling is not efficient enough to remove excess heat on dynamically short time scales,  
turbulence produced in clusters and filaments evolves somewhere in between the isothermal and adiabatic limits.

A key problem is how to identify turbulence in cosmological simulations and how to quantify its properties. The aforementioned studies come up with a remarkable variety of different approaches. \citet{NorBry99} introduced the standard deviation of the radial velocity in spherical shells around the center of a cluster as a proxy for turbulent fluctuations. This method has
been used in numerous studies ever since. However, it ignores transversal velocity fluctuations and obviously fails if the gas distribution is not spherically symmetric, for instance, in merging clusters or in filaments. To separate random turbulent fluctuations in three dimensions from large-scale bulk motions caused by the infall of gas into potential wells, \citet{DolVaz05} proposed to calculate a mean velocity field by smoothing the velocity on a scale of a few $100\;\mathrm{kpc}$. However, since turbulence produced by cosmic structure formation is neither statistically stationary nor homogeneous, it is not sufficient to apply a universal smoothing scale for the calculation of the mean flow. This limitation led \citet{VazzaRoed12} to the development of a filtering technique, which locally determines the coherence scale of the velocity field. The turbulent fluctuation is then obtained by subtracting the coherent component of the velocity. An alternative approach was initated by \citet{MaierIap09}, who applied a subgrid scale (SGS) model to determine the numerically unresolved fraction of the turbulent kinetic energy. However, their model entirely neglected effects caused by inhomogeneities and the shear associated with bulk flows. Apart from that, the SGS turbulence energy varies with the grid resolution. For this reason, it is not possible to infer the turbulent
velocity dispersion integrated over the whole spectrum of fluctuations. This is also the case if the vorticity of the velocity field is used as indicator of turbulence, which is meaningful only if the vorticity is calculated for a uniform grid resolution scale 
\citep[see, for example,][]{ZhuFeng10,IapSchm11,Miniati2014}. \citet{Miniati2014,Miniati2015} decomposes the velocity field into solenoidal and dilatational components and computes structure functions to infer scaling laws for turbulent velocity fluctuations. Particularly on small scales, the  spatial velocity increments should be dominated by turbulence. The integral-scale velocity fluctuation following from second-order structure functions provides a proxy for the turbulent velocity dispersion, with the caveat that bulk flows are not subtracted.

In this article, we apply the shear-improved SGS model implemented into the \nyx\
code \citep{AlmBell12,SchmAlm14} to compute the turbulent velocity dispersion in clusters and filaments. This model is applicable to inhomogeneous turbulence and provides a measure of turbulence with largely reduced dependence on numerical resolution and biases due to bulk flows. As explained in Section~\ref{sc:shock-gener-turb}, an adaptive algorithm originating from the statistical theory of signal processing is utilized to compute the mean velocity field. Similar to the multi-scale filtering method of \citet{VazzaRoed12}, the determination of the mean velocity is not completely unambiguous and has to be tweaked by sensible parameter choices.
While they employ filtering only as a postprocessing method, the shear-improved SGS model operates dynamically. This allows us to incorporate the turbulent stresses produced by numerically unresolved turbulent eddies into the equations of gas dynamics. Moreover, artificial changes in the internal energy induced by adaptive mesh refinement (AMR) are reduced.

The goal of this paper is to compute and to analyse statistics of the thermal and turbulent energies of the gas in a large sample of clusters, using the methodology outlined above and explained in more detail in Section~\ref{sc:num_methos}. In particular, we address the question if the turbulent Mach number tends to be constant. \citet{VazzaBrun11} proposed a ratio of $0.27$ between turbulent and thermal energies, corresponding to a Mach number of about $0.7$ for merger-driven turbulence in cluster cores). By computing second-order structure functions for massive clusters, \citet{MiniBer15} also found a nearly constant turbulent Mach number in the ICM. Moreover, we investigate whether turbulence approaches a statistically stationary state (balance of production and dissipation, see \citealt{MaierIap09}) and whether there are systematic differences between the turbulent energy in the WHIM and ICM \citep{IapSchm11}. The improved treatment of cooling processes and the UV background implemented into  \nyx\ by \citet{LukStark15} provides the basis for the accurate computation of the thermodynamic state resulting from structure formation. However, since we focus on turbulence driven by accretion and mergers at low redshift, we do not incorporate feedback processes. The impact of feedback on top of turbulence produced by structure formation will be studied in a separate work. 
After describing our simulation runs in Section~\ref{sc:runs}, we carry out a global analysis of gas properties (Section~\ref{sc:global}), followed by statistics of the mean turbulent velocity dispersion and temperature of the gas associated with halos of at least $10^{13}$ solar masses (Section~\ref{sc:halos}). To get a handle on resolution dependence and variations between different clusters, we also computed radial profiles (Section~\ref{sc:profiles}). The implications of our results and open questions are discussed in Section~\ref{sc:concl}.

\section{Shock-generated turbulence}
\label{sc:shock-gener-turb}

In the absence of galactic winds or AGN feedback, the dominant source
of turbulence in the intergalactic medium (IGM) or ICM is the interaction of shocks produced during and after the nonlinear
gravitational collapse of baryons \citep[e.~g.][]{KangRyu07,RyuKang08,VazzaBrun09b,PaulIap11,Miniati2014}. 
Particularly at lower redshifts, gas falling into the deep potential wells of clusters is accelerated to 
high peculiar velocities $U_{\rm u}\sim 1000\,\mathrm{km/s}$. Since the temperature $T_{\rm u}$ 
of the low-density IGM from which gas is accreted does not exceed a few $10^4\,\mathrm{K}$, such velocities correspond to hypersonic upstream Mach numbers $\mathcal{M}_{\rm u} \gg 1$.
As a result, very strong shocks are produced \citep{MiniRyu00,RyuKang03,SkillShea08}. The Mach number $\mathcal{M}_{\rm s}$ with respect to the shocked gas follows from the jump conditions for a strong shock:
\begin{equation}
  \label{eq:mach1}
  \mathcal{M}_{\rm s} \simeq \left(\frac{\gamma-1}{2\gamma}\right)^{1/2}\,, 
\end{equation}
giving $\mathcal{M}_{\rm s} \sim 0.45$ for $\gamma = 5/3$. The shock-compressed gas
is the main constituent of the WHIM. Since accretion shocks are located far outside the center of a cluster, the gas can still gain kinetic energy by being pulled toward denser regions of the cluster. Since the gas is also turbulent, however, kinetic energy is efficiently dissipated
into thermal energy. For this reason, the Mach number does not necessarily become higher. If radiative cooling is taken into account, the dynamics becomes even more complex because
now turbulent heating competes with cooling, which may raise the Mach number.

Apart from accretion, the gas in clusters is stirred by mergers. 
For the subclass of merger shocks produced by the motion of substructure halos within large virialized halos
(see, for example, \citealt{Cen2005,SubShu06,IapAdam08,RussFab14}), 
a simple estimate of the turbulent Mach number can be obtained by noticing that both the typical velocity of the subhalo $v_{\rm v}$ and the sound speed in the gas of the larger halo $c_{\rm s}$ are given by the
larger halo's virial temperature \citep[see also][]{FaltKarvt05}, 
\begin{equation}
  \label{eq:mach2}
  T_{\rm v} \sim \frac{\mu m_{\rm p} v_{\rm v}^2}{2 k_{\rm B}} \sim \frac{\mu
    m_{\rm p} c_{\rm s}^2}{\gamma k_{\rm B}} \,\,,
\end{equation}
implying a weak shock with upstream Mach number $\mathcal{M}_{\rm u} \sim (2/\gamma)^{1/2} \approx 1.1$ for $\gamma = 5/3$. The post-shock Mach number $\mathcal{M}_{\rm s}$ is also close to unity in this case. A more general analysis of merger shocks, implying Mach numbers of order unity, is presented in \citet{GabBlasi03}. In contrast to mergers of clusters with comparable mass, numerical simulations of minor mergers show that turbulence is mainly produced in the turbulent wake of the subhalo \citep{IapAdam08,SchmSchulz15,RoedKraft15}, resulting in subsonic turbulent Mach numbers. This is also suggested by observations of \citet{EckMol14}. 

Based on simple dimensional arguments, one can therefore expect that shock-generated turbulence is roughly transonic.
However, local differences in the efficiency of transforming downstream shock velocities into turbulence and additional kinetic energy injection by gravity (i.~e., acceleration toward the center of the gravitational well inside the outer shocks and stirring through mergers) potentially vary the turbulent energy, while the thermal energy of the gas changes due to turbulent energy dissipation, gravitational compression, and radiative cooling. The combined effect of these processes can only be studied with numerical simulations and will be reflected in a large scatter of Mach numbers. 

\section{Numerical methods}
\label{sc:num_methos}

We perform cosmological simulations with the cosmological $N$-body and fluid dynamics
code \nyx\ \citep{AlmBell12}, which has been developed at the Lawrence Berkeley National Laboratory. 
The code is based on the \textsc{BoxLib} framework for block-structured adaptive mesh refinement, which follows the procedure outlined in \citet{bell-3d}, with a multi-grid solver
for the computation of the gravitational potential and optimal subcycling for the adaptive hydrodynamical time step \citep{AlmBell12}.
The positions and velocities of the dark matter particles are evolved using a kick-drift-kick sequence such as in \citet{miniati-colella}. 
To solve the equations of gas dynamics in co-moving coordinates, an unsplit Godunov method with piecewise parabolic reconstruction is applied \citep{miller-colella}.\footnote{\citet{SchmSchulz15} compare the \nyx\ solver
	to a standard PPM implementation with directional splitting for a minor-merger scenario.}
For a more realistic treatment of the baryonic gas,  we use an advanced treatment of heating and cooling on cosmological scales, as detailed in \citet{LukStark15}. However, we do not consider contributions from feedback processes or magnetic fields here.

In cosmological simulations, it is commonly assumed that a fluid-dynamical description is applicable to the baryonic gas at least on sufficiently large scales. Moreover, the physical dissipation scale due to microscopic viscosity is assumed to be unresolved in such simulations. In contrast to implicit large eddy simulations (ILES), where the effect of turbulent diffusion and energy dissipation is solely treated by the truncation error terms of the finite-volume discretization, large eddy simulations (LES) with an explicit subgrid scale (SGS) model approximate the kinetic energy of numerically unresolved turbulence by an additional transport equation (see \citealt{Schm15} for a review). The additional energy variable, which is called SGS turbulence energy, constitutes a buffer between the numerically resolved kinetic energy and the internal energy of the gas.
For this reason, it is also possible to apply an improved, energy- and momentum-conserving AMR scheme, where artificial modulations of the internal energy due to grid refinement or de-refinement are reduced \citep[see][]{SchmAlm14}.
Since the turbulent diffusivity predicted by the SGS model allows us to account for the
enhanced transport of heat due to small-scale turbulence, we incorporated a turbulent diffusion term into the equation for the numerically resolved energy. Thus, \nyx\  solves the following equations for the comoving baryonic density $\rho$, the proper peculiar velocity $\Ub$ (i.~e., the physical velocity minus the Hubble flow), the numerically resolved energy $\rho E = \rho(e + \Ub \cdot \Ub/2)$, where $\rho e = p/(\gamma - 1)$ is the internal energy of gas with pressure $p$ and adiabatic exponent $\gamma$,
and the SGS turbulence energy $\rho K$ \citep{SchmAlm14}:
\begin{align}
	\label{eq:mass_les}
	\frac{\partial \rho}{\partial t} =& - \frac{1}{a} \nabla \cdot (\rho \Ub) \, , \\
	\label{eq:momt_les}
	\frac{\partial (a \rho \Ub)}{\partial t} =& 
	-             \nabla \cdot (\rho \Ub\otimes\Ub) 
	-             \nabla p
	+             \nabla \cdot \boldsymbol{\tau}
	+             \rho \gb  \, ,\\
	\label{eq:energy_int_les}
	\begin{split}
	\frac{\partial (a^2 \rho e)}{\partial t} =&
	 - a \nabla \cdot (\rho \Ub e) - a p \nabla \cdot \Ub
	+ a \dot{a} [2 - 3 (\gamma - 1) ] \rho e\\
	&+ a\sub{\Lambda}{HC}
	+a \nabla \cdot (\rho \kappa_{\rm sgs}\nabla e)+ a\rho \varepsilon\, ,
	\end{split}
\end{align}
\begin{align}
	\label{eq:energy_les}
	\begin{split}
	\frac{\partial (a^2 \rho E)}{\partial t} =& - a \nabla \cdot (\rho \Ub E + p \Ub)
	 + a (\rho \Ub \cdot \gb + \sub{\Lambda}{HC})\\
	&+ a \dot{a} [2 - 3 (\gamma - 1)] \rho e
	+ a \nabla \cdot (\Ub\cdot\boldsymbol{\tau} + \rho \kappa_{\rm sgs}\nabla e)\\
	&- a(\Sigma - \rho \varepsilon) \, ,	
	\end{split}\\
	\label{eq:k_les}
	\begin{split}
	\frac{\partial (a^2\rho K)}{\partial t} =&
	 - a\nabla \cdot \left(\rho \Ub K\right) 
	 + a\nabla \cdot \left(\rho \kappa_{\rm sgs}\nabla K\right) \\
	&+ a(\Sigma - \rho \varepsilon)\, ,
	\end{split}
\end{align}
The gravitational acceleration vector in equations~(\ref{eq:momt_les}) and~(\ref{eq:energy_les}) is given by $\gb = - \nabla \phi$, where $\phi$ is the peculiar potential resulting from the total density field of baryonic and dark matter. The combined rate of heating and cooling, $\sub{\Lambda}{HC}$, in equations~(\ref{eq:energy_int_les}) and~(\ref{eq:energy_les}) is computed for the time-varying homogeneous ultraviolet background from \citet{HaardtMad12} and radiative cooling due to hydrogen and helium \citep{KatzWein96,LukStark15}. The rate of SGS turbulence energy production in equations~(\ref{eq:energy_les}) and~(\ref{eq:k_les}) 
is defined by
\begin{equation}
	\label{eq:prod}
	\Sigma = \tau_{ij} S_{ij} \, ,
\end{equation}
and 
\begin{equation}
	\label{eq:diss}
	\varepsilon = \frac{C_\varepsilon K^{3/2}}{\Delta}
\end{equation}
is the the dissipation rate.

The non-linear interaction between numerically resolved and unresolved scales is given by the SGS turbulence stress tensor $\boldsymbol{\tau}$. To compute $\boldsymbol{\tau}$ from known variables, we use the following non-linear closure for compressible turbulence 
\citep{SchmFeder11}:
\begin{equation}
 \label{eq:tau_nonlin}
 \begin{split}
  \tau_{ij} =&\, 2C_{1}\Delta\rho(2 K_{\mathrm{sgs}})^{1/2}S_{\! ij}^{\ast}
   -4C_{2}\rho K\frac{U_{i,k}U_{j,k}}{|\nabla\otimes\Ub|^{2}}\\
  &-\frac{2}{3}(1-C_{2})\rho K\delta_{ij}
  \end{split}
\end{equation}
with constant coefficients $C_1 = 0.02$ and $C_2 = 0.7$, the grid scale in comoving coordinates, $\Delta$, the Jacobian matrix $U_{i,k}$ of the resolved velocity, its norm $|\nabla\otimes\Ub|:=(2U_{i,k}U_{i,k})^{1/2}$, and the trace-free rate-of-strain tensor
\begin{equation}
	\label{eq:strain}
	S_{ij}^{\ast} = S_{ij} - \frac{1}{3}\delta_{ij}d \,:=
	\frac{1}{2} (U_{i,j} + U_{j,i}) - \frac{1}{3}\delta_{ij}U_{k,k}\,.
\end{equation}
The quantity $d=U_{k,k}$ is called the divergence of the velocity field.

To separate turbulent velocity fluctuations from the non-turbulent bulk motions, an adaptive Kalman filter is introduced in \citet{SchmAlm14}. While bulk motions are, for instance, associated with large-scale accretion flows into the potential wells of clusters, turbulence is produced by accretion shocks and merger events. This separation leads to the shear-improved SGS model \citep{CahuBou10}, which allows us to calculate the turbulent velocity dispersion for clusters without strong biases from bulk flows:
\begin{equation}
  \label{eq:sigma_turb}
   \sigma_{\rm turb}^2 = U^{\prime\,2} + 2K,
\end{equation}
where $\Ub^\prime=\Ub-[\Ub]$ represents the fluctuating component of the numerically resolved flow. The smoothed flow velocity $[\Ub]$ 
is locally computed with an adaptive temporal Kalman filter. 
In contrast to a simple low-pass filter, such as the exponential filter, this method has the advantage that no fixed smoothing scale is applied. It is therefore particularly suitable for inhomogeneous turbulent flow. However, the adaptive algorithm requires guide parameters, which control the behaviour of the filter if the fluctuating field settles into equilibrium. In the statistically stationary limit, the filter operates like an exponential low-pass filter with the characteristic time scale $\sub{T}{c}=5\;\mathrm{Gyr}$.\footnote{
	See \citet{SchmAlm14} for a detailed explanation of the calibration of the Kalman filter in
	the case of the Santa Barbara cluster. Here, we use the parameters $\sub{T}{c}=5\;\mathrm{Gyr}$
	and $\sub{U}{c}=300\;\mathrm{km/s}$ for the predictor-corrector scheme. Compared to the Santa
	Barbara cluster, we lowered $\sub{U}{c}$ by $25\,\%$ to work with a value that is representative
	for the turbulent velocity dispersion over a wide range of overdensities (see the global phase
	plot for $\sigma_{\rm turb}$ vs.\ overdensity in Sect.~\ref{sc:global}). The time scale 
	$\sub{T}{c}$, on the other hand, turned out to be very robust. It is important to notice
	that $\sub{U}{c}$ is not simply a characteristic scale of the fluctuating component 
	(if that were the case, the separation would be rather arbitrary). In fact, 
 	a higher value of $\sub{U}{c}$ results in a \emph{lower} magnitude of $\Ub^\prime$. 
 	However, if $\sub{U}{c}$ is appropriately chosen, the resulting turbulent velocity dispersion
 	should be roughly comparable to $\sub{U}{c}$ in the statistically stationary regime (see \citealt{SchmAlm14,Schm15} for further details).}
By invoking ergodicity, spatial scales could be associated with the time scale of the filter. However, depending on the variability of the flow, the resulting length scales can range from the integral scale in the case of homogeneous and isotropic turbulence to much smaller scales for rapid spatiotemporal changes in the flow, for example, across accretion shocks.

The turbulent velocity dispersion predicted by our model encompasses both resolved and
subgrid scales. Moreover, it is possible to calculate the timescale of turbulence production
associated with energy transfer through the turbulent cascade:
\begin{equation}
  \label{eq:tau_dyn}
  \tau = \frac{\rho\sigma_{\rm turb}^2}{2\Sigma}\,.
\end{equation}

For the simple eddy-viscosity closure applied in \citet{SchmAlm14}, the application of the shear-improved model is straightforward. However, a subtlety arises when working with the mixed closure~(\ref{eq:tau_nonlin}) for the turbulent stresses.
\citet{LevTosch07} argue that the turbulence energy flux $\Sigma$ should be linear in the 
derivative $U_{i,j}^\prime$ of the fluctuating component. While this is naturally the case for the eddy-viscosity term with the coefficient $C_1$, replacing $U_{i,j}$ by $U_{i,j}^\prime$ in the non-linear term with the coefficient $C_2$ would result in a quadratic dependence on the fluctuating component. We remove this inconsistency by observing that the main function of the shear-improved treatment is to suppress turbulence production in regions where shear is mainly due to the bulk flow. In this case, $K$ should be small and the dominant term is the first term in the closure for $\tau_{ij}$.
The non-linear term, on the other hand, is only important in strongly turbulent regions \citep[see][]{SchmFeder11}, where locally isotropic developed turbulence can be assumed and  the shear-improved correction is small. For this reason, we defined the shear-improved SGS turbulence stress tensor as
\begin{equation}
  \label{eq:tau_si}
  \begin{split}
	\tau_{ij}^{\rm (SI)}=&\, 2C_{1}\Delta\rho(2 K_{\mathrm{sgs}})^{1/2}
	\left[\frac{1}{2} (U_{i,j}^\prime + U_{j,i}^\prime)
	      -\frac{1}{3}\delta_{ij}U_{k,k}^\prime\right] \\
	&-4C_{2}\rho K\frac{U_{i,k}U_{j,k}}{|\nabla\otimes\Ub|^{2}} - \frac{2}{3}(1-C_{2})\rho K\delta_{ij}\,.
  \end{split}
\end{equation}

\section{Simulation runs}
\label{sc:runs}

We use the \textsc{Music} code of \citet{HahnAbel11} to produce cosmological 
initial conditions in a box of about $152\;\mathrm{Mpc}$ comoving size, which 
is sufficiently large to contain several big clusters and major mergers (see Section~\ref{sc:global}). 
Following the results of the \citet{Planck14}, the chosen cosmological density parameters are $\Omega_{\rm m}=0.315$ for matter,
$\Omega_{\rm b}=0.0487$ for the baryonic matter component, and $\Omega_\Lambda=0.685$ for the cosmological constant.
The Hubble parameter at $z=0$ is $H_0 = 67.3\;\mathrm{(km/s)/Mpc}$. 

An overview of the runs performed with different numerical resolutions is given in Table~\ref{tb:runs}. 
We varied both the grid resolution and the number of dark matter particles and applied refinement criteria based 
on overdensity and vorticity as described in \cite{SchmAlm14}.\footnote{However, 
an overdensity factor of at least $8$ must be reached in the current 
simulations, which implies that each of the new cells at the next higher refinement level contains the same mass as the initial mass in the coarser cell. 
Refinement by vorticity ensures that turbulent regions are covered by the 
highest refinement level see \citealt{IapAdam08,VazzaBrun11,Miniati2014} for alternative approaches}). Since turbulent gas 
fills vast regions of space, which are bounded by the accretion shocks 
surrounding clusters and thick filaments, the filling factors of refined cells can grow above $60\,\%$ at the first refinement level and to nearly $20\,\%$ at the second level.  This is why our simulations tend to be computationally quite expensive.
To appreciate the size of the simulations, the effective resolution and the 
filling factors at redshift $z=0$ are also listed in Table~\ref{tb:runs}. Of 
course, higher spatial resolution would be necessary if feedback processes from 
galaxies were included. With a comparable computational budget, this can be 
achieved by reducing the box size, applying less aggressive refinement 
strategies or zooming into selected clusters. In this article, however, we focus 
on the statistics of a large sample of big clusters at the high-mass end.

\begin{table}
  \begin{center}
      \begin{tabular}{c|rcrrccc}
        \hline
        Run & $N_0=N_{\rm p}$ & $\sub{l}{max}$ & $\Delta_{\rm min}$ & \sub{N}{eff} & 
        $V_1/V_0$ & $V_2/V_0$ \\ 
       \hline\hline
       $\text{C}_1$ & $256^3$  & $1$ & $\unit{296}{kpc}$ & $512^3$  & 0.14 &  \\
       $\text{C}_2$ & $256^3$  & $2$ & $\unit{148}{kpc}$ & $1024^3$ & 0.59 & 0.19 \\
       $\text{B}_1$ & $512^3$  & $1$ & $\unit{148}{kpc}$ & $1024^3$ & 0.32 &      \\
       $\text{B}_2$ & $512^3$  & $2$ & $\unit{74}{kpc}$  & $2048^3$ & 0.64 & 0.17 \\
       $\text{A}_1$ & $1024^3$ & $1$ & $\unit{74}{kpc}$  & $2048^3$ & 0.37 &      \\
       \hline                
      \end{tabular}
  \end{center}
  \caption{Overview of simulation runs, ordered by increasing numerical resolution, where 
  	$N_0$ is the number of root-grid cells, $N_{\rm p}$ the number of dark matter particles, 
	$\sub{l}{max}$ the maximum refinement level,
	$\Delta_{\rm min}$ is the spatial resolution at the highest refinement level,
	and $\sub{N}{eff}=2^{3\sub{l}{max}}N_0$ the effective resolution. 
	The volume filling factors of levels 1 and 2 at the end of a simulation 
	($z=0$) are	$V_1/V_0$ and $V_2/V_0$, respectively.}
  \label{tb:runs}
\end{table}

\begin{figure*}
\centering
\includegraphics[width=\linewidth]{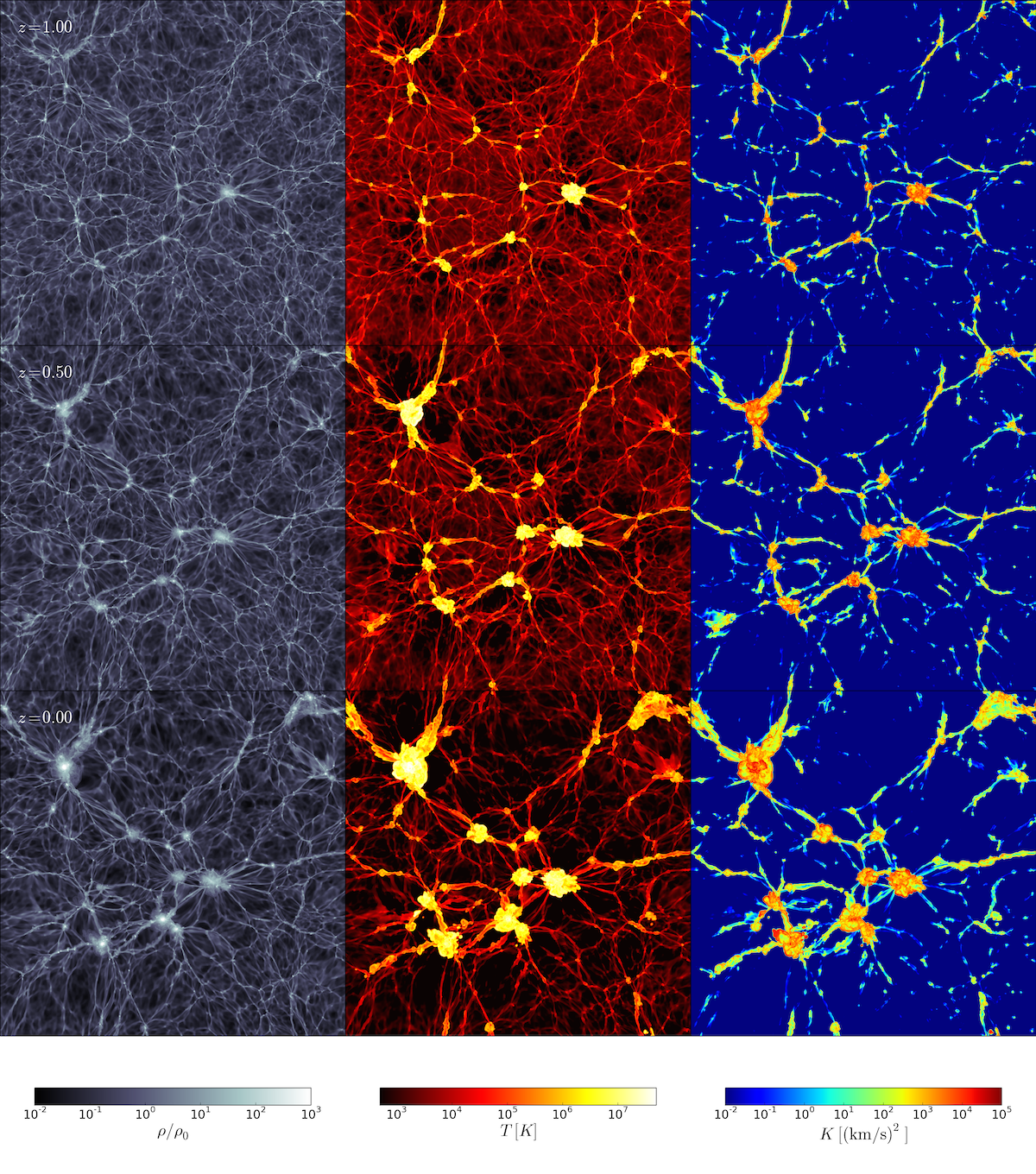}
\caption{Slices of the gas overdensity $\rho/\rho_0$ (left), the temperature $T$ (middle), and the 	 	specific SGS turbulence energy $K$ (right) at three different redshifts for run $\text{A}_1$.}
\label{fig:slices_evol}
\end{figure*}

\begin{figure*}
\centering
\includegraphics[width=\linewidth]{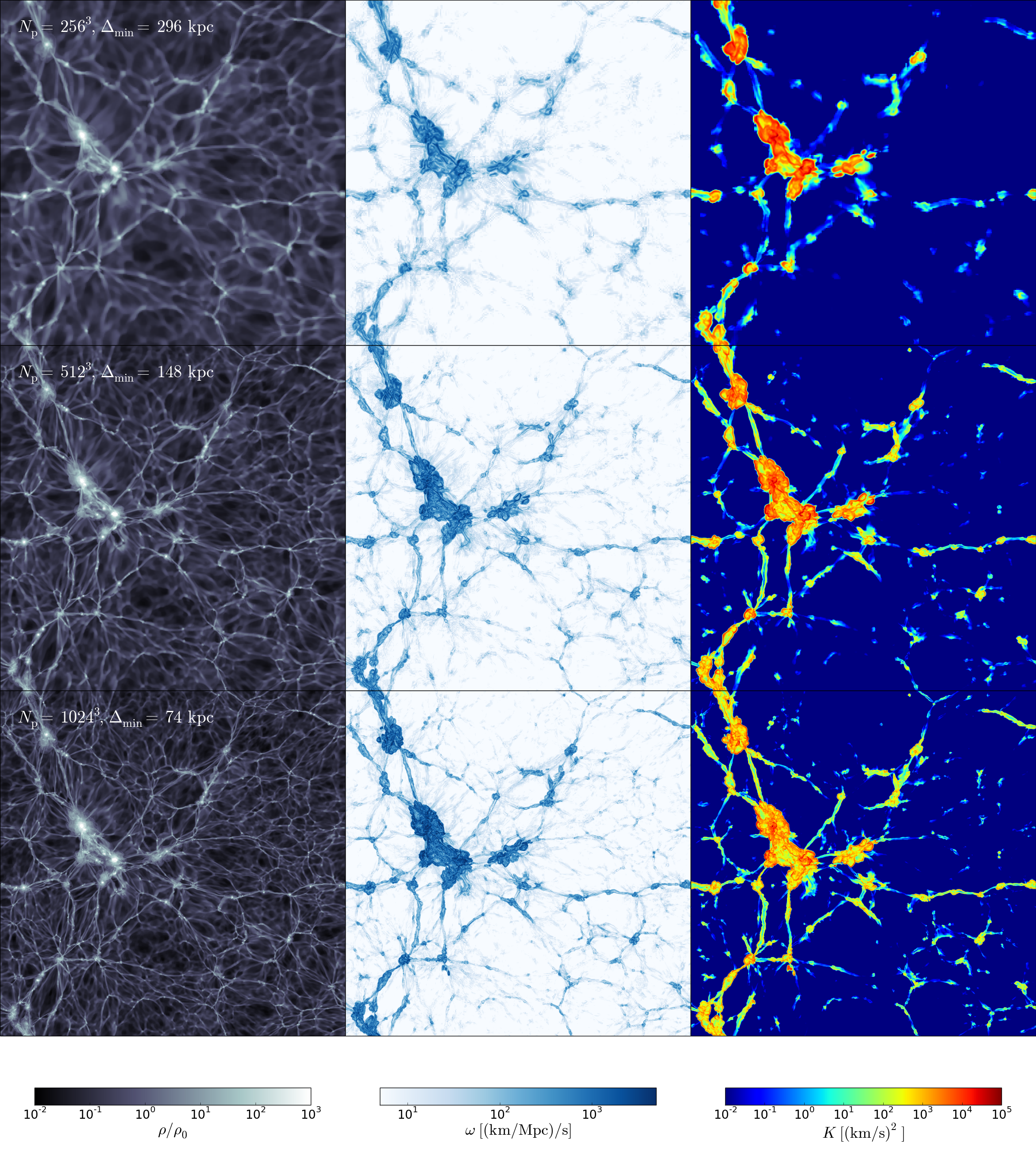}
\caption{Slices of the gas overdensity $\rho/\rho_0$ (left), the vorticity 
$\omega$ (middle), and the specific SGS turbulence energy $K$ 
	(right) 
	at redshift $z=0$ for runs $\text{C}_1$ (top), $\text{B}_1$ (middle), and $\text{A}_1$ (bottom)
	with different particle numbers $N_{p}$ and numerical resolution scales $\Delta$.}
\label{fig:slices_res}
\end{figure*}

Figure~\ref{fig:slices_evol} shows slices of the gas overdensity, the temperature, and the
SGS turbulence energy per unit mass for the highest-resolution run in a different plane. At $z=0$ (bottom panels), a network of several big clusters connected by filaments can be seen. Close to the upper left corner of the plane, a particularly massive and strongly turbulent clusters is produced by a major merger between $z\approx 1.0$ (upper panels) and  $0.5$ (middle panels). The temperature in the ICM of the big clusters increases to several $10^7\;\mathrm{K}$. The accretion shocks in the cluster outskirts are indicated by a sharp drop of the temperature. The slices of the specific SGS turbulence energy show that significant turbulence develops only inside the accretion shocks. Similar to our previous simulations of 
the Santa Barbara cluster \citep{SchmAlm14}, the shear-improved SGS model avoids spurious SGS turbulence energy production in the low-density gas outside the accretion shocks, which show no significant turbulence on numerically resolved length scales.

The influence of numerical resolution is illustrated by slices of the gas overdensity, the vorticity modulus, and the SGS turbulence energy per unit mass in Fig.~\ref{fig:slices_res}. 
We chose a different section through the box, showing an enormous complex of clusters at $z=0$. These clusters are about to merge. Although the impact of the grid and particle resolution on the structure of the cosmic web is palpable, the vorticity 
$\bmath{\omega}=\vecnab\times\vecU$ indicates turbulent gas in clusters for all resolutions. If the clusters are sufficiently massive so that the regions inside accretion shocks extend over many resolution elements, turbulence is also indicated by high values of the SGS energy $K$. For these clusters, one can see that $K$ tends to increase if the resolution is lowered because an increasing fraction of turbulence energy resides in unresolved modes. However, for small clusters or isolated galaxies, which are only marginally resolved, little or no SGS energy can be seen in the lowest-resolution run (top left slide in Fig.~\ref{fig:slices_res}). This is indicative of insufficient resolution: If the dynamical range of numerically resolved modes is too narrow, fluctuations with respect to the mean flow calculated with the Kalman filter are strongly damped by the truncation error of the
numerical scheme. In this case, the leading-order term in equation~(\ref{eq:tau_si}) and, consequently, the growth of $K$ are suppressed. We will further elaborate on resolution requirements on the basis of quantitative measures in Sections~\ref{sc:halos} and~\ref{sc:profiles}.

\begin{figure*}
\centering
    \includegraphics[width=0.49\linewidth]{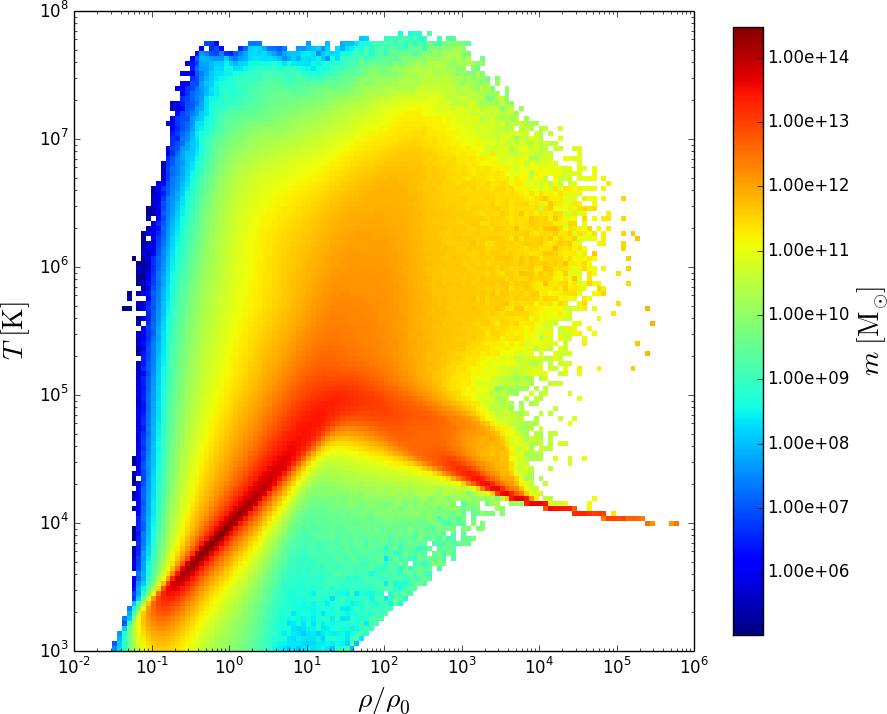}\quad
    \includegraphics[width=0.49\linewidth]{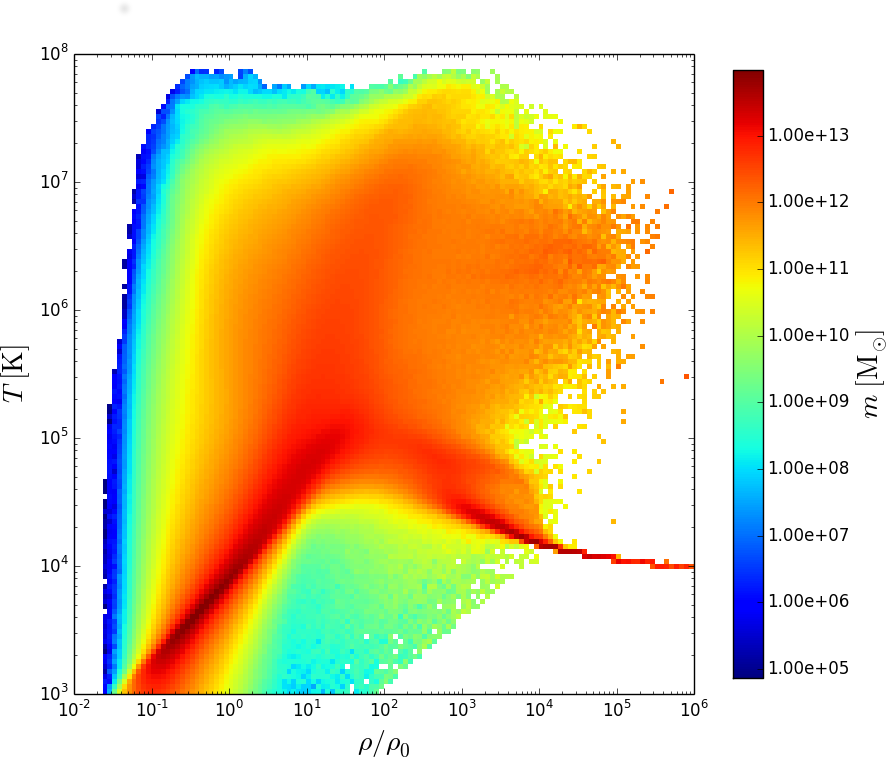}
    \includegraphics[width=0.49\linewidth]{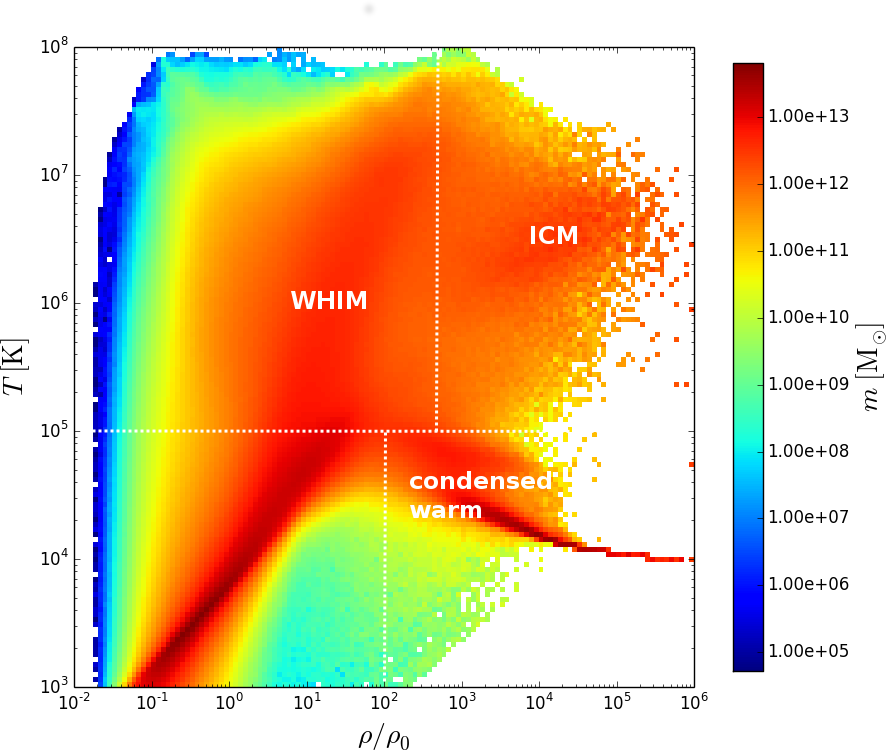}\quad
    \includegraphics[width=0.49\linewidth]{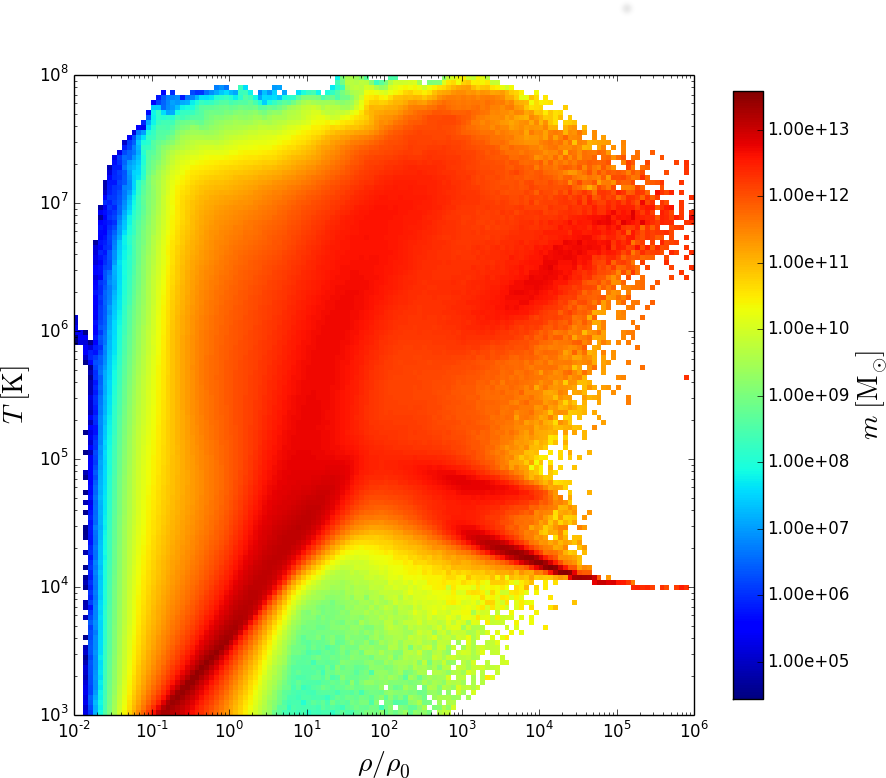}
    \caption{Phase diagrams (temperature vs.\ overdensity) for run $\text{B}_2$ at redshifts 
    $z=2.0$ (top left), $1.0$ (top right), $0.5$ (bottom left), and $0.0$ (bottom right).
    For $z=0.5$, the approximate separation into different phases is indicated by
	the white dotted lines.}
    \label{fig:phase}
\end{figure*}

\begin{figure*}
\centering
    \includegraphics[width=0.49\linewidth]{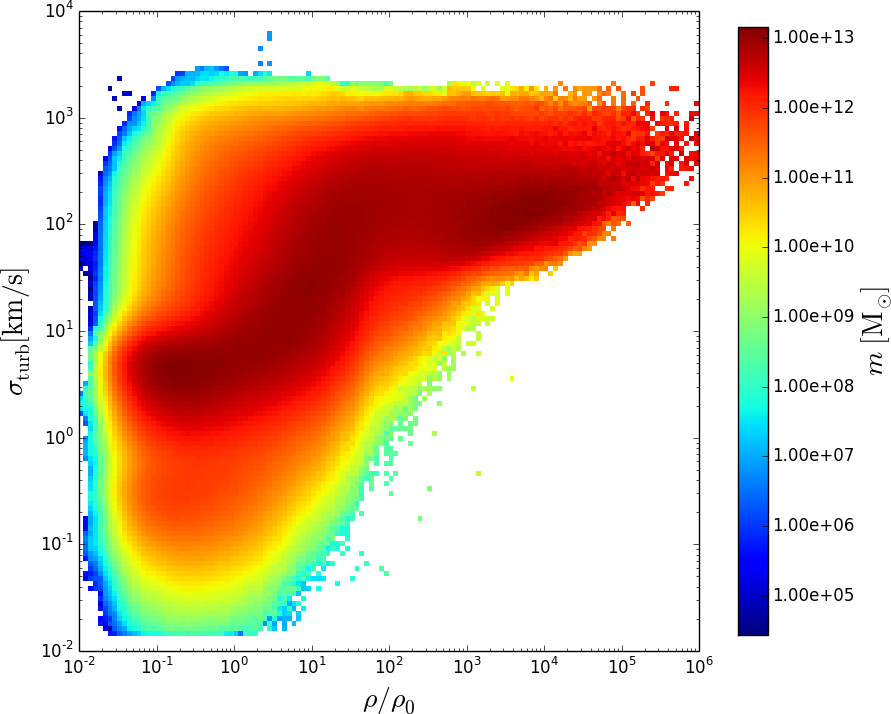}\quad
    \includegraphics[width=0.49\linewidth]{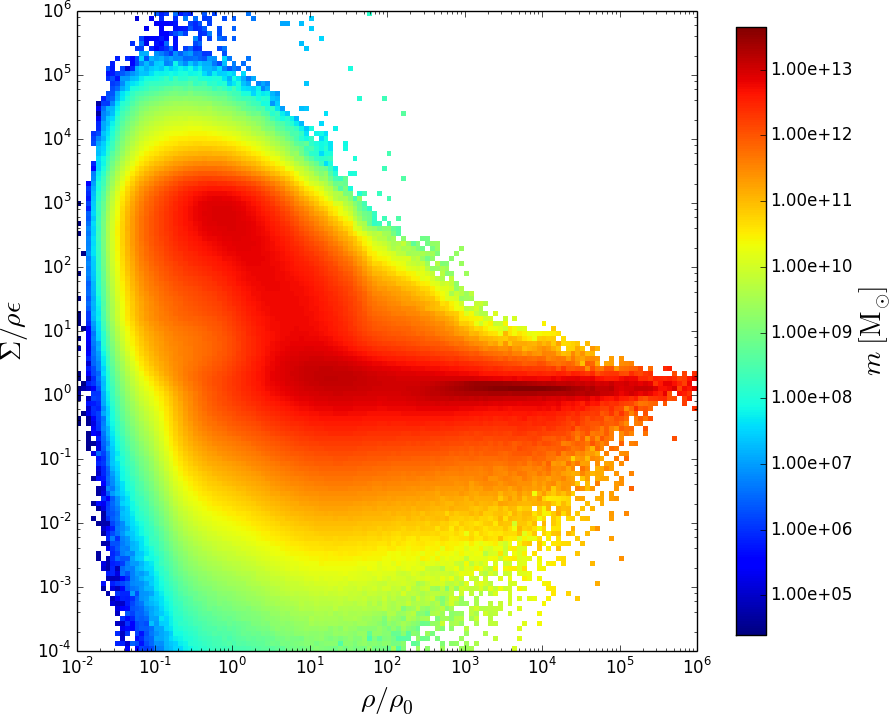}
    \caption{Turbulent velocity dispersion (left) and production-to-dissipation ratio (right) vs.\  
    	overdensity for run $\text{B}_2$ at redshift $z=0.0$.}
    \label{fig:phase_turb}
\end{figure*}

\section{Global statistics}
\label{sc:global}

In Figure~\ref{fig:phase}, the heating of the gas due to mergers and shock compression can be seen in phase diagrams showing the gas mass in logarithmic bins of the baryonic overdensity $\rho/\rho_0$ and the temperature $T$ for redshifts ranging from $z=2.0$ to $0.0$. At $z=2.0$, most of the gas can be found in two branches. 
On the one hand, the rising polytrope for densities around the mean density stems from the UV background resulting from the reionization model of \citet{HaardtMad12}. At high densities, on the other hand, the gas can condense into an ISM-like phase of warm gas \citep{KatzWein96,LukStark15}. Since we do not account for cooling processes in the low-temperature phases of the ISM (galaxies are not resolved in our simulations), the high-density branch asymptotically approaches $10^4\,\mathrm{K}$. As cosmological structure formation progresses, both branches are still visible, but an increasing amount of gas is heated to temperatures much higher than $10^5\;\mathrm{K}.$ For $z=0.5$ (bottom left plot in Fig.~\ref{fig:phase}), an accumulation of gas with overdensities above $\sim 10^3$ and temperatures in the range from about $10^6$ to several $10^7\;\mathrm{K}$ is discernible, which corresponds to the ICM of massive clusters. The gas at lower densities ($\rho/\rho_0\lesssim 100$) with temperatures above $10^5\;\mathrm{K}$ constitutes the warm-hot intergalactic medium (WHIM; see also \citealt{IapSchm11,SmithHall11}). The temperature threshold $T=\unit{10^5}{K}$ constrains the volume to regions inside the accretion shocks and thick filaments, while the gas in the void and the gas that condenses into the low-temperature and high-density branch is excluded. In the diagram for $z=0.5$ (bottom left plot), the temperature and density bounds used in the following are indicated by the white dashed lines. The density threshold $\rho/\rho_0 = 500$ was chosen to separate WHIM and ICM roughly along the depression between the two maxima of the gas mass for temperatures $T\sim 10^6\;\mathrm{K}$. In contrast to \citet{SmithHall11}, we do not impose an upper temperature bound of $10^7\;\mathrm{K}$ because a separate hot phase
is not motivated by the gas distribution that can be seen in the phase diagrams. We rather see an increasing amount of gas gradually spreading out over temperatures up to $10^8\;\mathrm{K}$ with decreasing redshift. The very hot gas resides mainly in the large clusters forming in our simulation. Moreover, we separate the ``warm'' gas below $T=\unit{10^5}{K}$ into a tenuous ($\rho/\rho_0 < 100$) and a condensed ($\rho/\rho_0 > 100$) phase to account for the major branches resulting from local equilibria of heating and cooling \citep[see also][]{LukStark15}. The gas phases are further analyzed for different halos in Section~\ref{sc:halos}.
 
Since Fig.~\ref{fig:slices_evol} suggests that mergers and accretion produce 
turbulence, hot overdense gas should be associated with strong turbulence. The 
distribution of the turbulent velocity dispersion $\sub{\sigma}{turb}$ defined 
by equation~(\ref{eq:sigma_turb}) is shown in Fig.~\ref{fig:phase_turb} (left 
plot). Indeed, one can see that most of the gas with $\rho/\rho_0\gtrsim 10$ is 
highly turbulent, with a typical velocity dispersion of few $100\;\mathrm{km/s}$. Around 
the mean density $\rho_0$, there is a sharp drop to $\sim 1\;\mathrm{km/s}$.  
The right plot in Fig.~\ref{fig:phase_turb} shows the ratio of the turbulence 
production rate $\Sigma$ (equation~\ref{eq:prod}) to the dissipation rate 
$\rho\epsilon$ (equation\ref{eq:diss}) against the overdensity. A pronounced peak 
of the gas mass can be seen at $\Sigma/(\rho\epsilon)\sim 1$ in the high-density 
range ($\rho/\rho_0\gtrsim 10^3$). This implies an equilibrium between 
production and dissipation in the ICM. In low-density gas, on the other hand, 
turbulence is far off equilibrium.  For a large fraction of gas around the mean 
density, $\Sigma/(\rho\epsilon)\gg 1$, i.~e., turbulence is being produced while 
dissipation is still inefficient. This is in agreement with the small turbulent 
velocity dispersion at low densities that can be seen in the left plot.  

\begin{figure}
\centering
    \includegraphics[width=\linewidth]{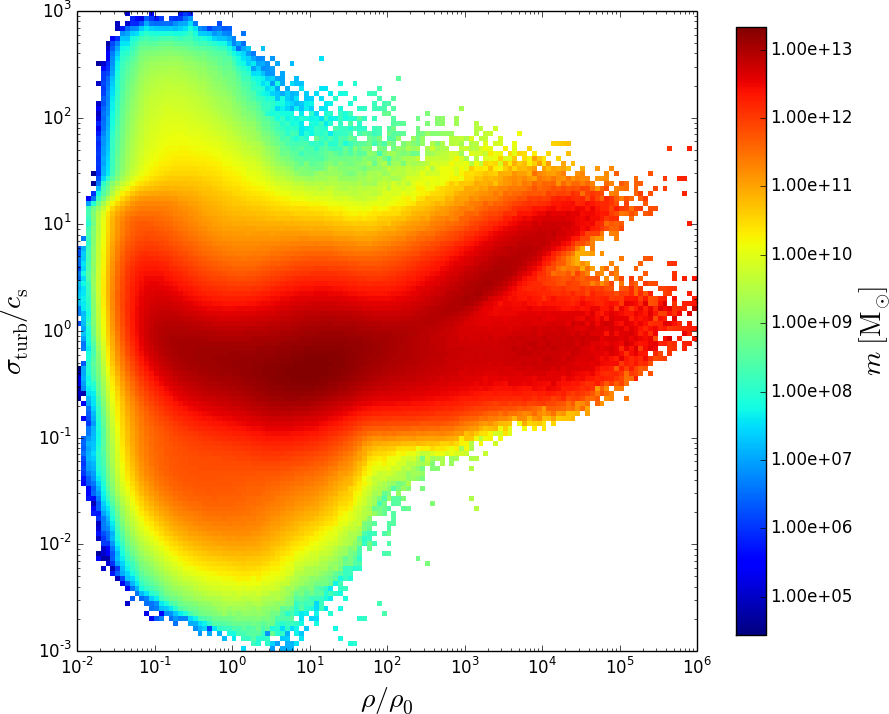}
    \caption{Mach number vs.\ overdensity for run $\text{B}_2$ at redshift $z=0.0$.}
    \label{fig:phase_mach}
\end{figure}

The Mach number $\sub{\sigma}{turb}/\sub{c}{s}$ associated with turbulence has a supersonic and a transonic branch at high densities (see Fig.~\ref{fig:phase_mach}). The difference of about $1.5$ orders of magnitude in the Mach numbers is roughly comparable to the three orders of magnitude in temperature ($T\propto\sub{c}{s}^2$), corresponding to the cold-gas branch and the hot ICM in the phase diagram. This implies a similar intensity of turbulence in both phases. At moderate overdensities (WHIM), Mach numbers around unity are dominant.     

\section{Halo statistics}
\label{sc:halos}

The relation between turbulence and temperature for clusters can be quantitatively demonstrated by calculating mean values in spherical regions associated with dark-matter halos. We employed the halo finder implemented in yt \citep{TurkSmith11} to identify these regions. For each halo, a sphere with radius $5\sub{R}{halo}$ centered at the local density peak was extracted, where $\sub{R}{halo}$ is 
the maximal radial distance of a dark-matter particle belonging to the halo from its center. The multiplication by $5$ accounts for the much larger volume enclosed by the outer shock fronts compared to the size of the halo, as defined by the halo finder. A factor of $5$ turned out to be a robust choice, which ensures that both the WHIM and the ICM are contained in the spherical region, for which averages are computed.

\begin{figure*}
\centering
    \includegraphics[width=\linewidth]{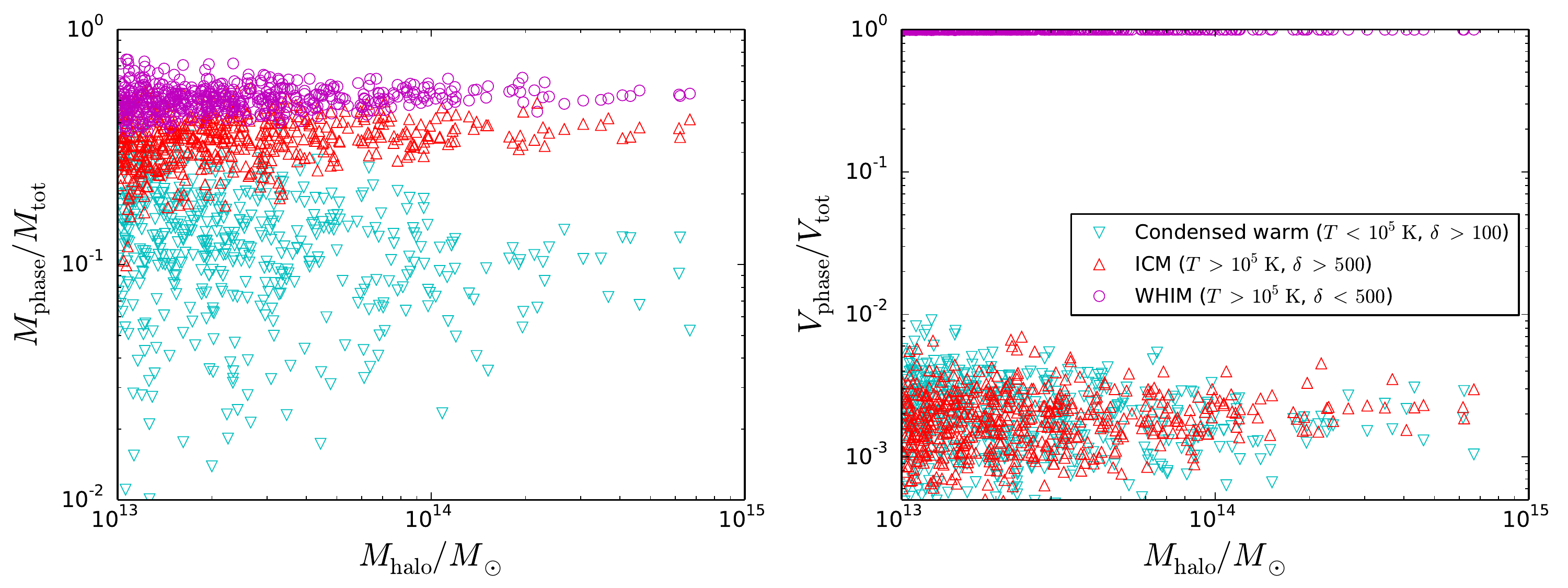}
    \caption{Mass and volume fractions of the different phases (WHIM, ICM, and condensed 
    warm gas) associated with halos of mass $M_{\rm halo}$ at $z=0$ (run $\text{A}_1$). 
    The total gas mass is defined by $M_{\rm tot}=M_{\rm WHIM}+M_{\rm ICM}+M_{\rm cw}$ and, 
    analogously, the total volume.}
    \label{fig:gas_frac}
\end{figure*}

The volume and mass fractions of the different phases defined in Section~\ref{sc:global} and indicated in the phase plot in Fig.~\ref{fig:phase} are plotted for all halos with dark-matter mass $M>10^{13}\,M_\odot$ at $z=0$ in Fig.~\ref{fig:gas_frac}.\footnote{Except for Fig.~\ref{fig:gas_frac}, where the subscript halo is used for clarity, it is understood in the following that $M$ 	refers to the halo mass, not the gas mass.}
The mass is normalized to the total mass of gas that has either temperatures above $10^{5}\;\mathrm{K}$ (WHIM and ICM) or an overdensity greater than $100$ (condensed warm gas). Excluded is the tenuous warm gas ($T<10^{5}\;\mathrm{K}$ and $\rho/\rho_0 < 100$). The WHIM and ICM clearly contain most of the gas mass, particularly for clusters with halo mass greater than $10^{14}\,M_\odot$. The
typical mass fraction (see also Table~\ref{tb:frac}) of the WHIM is about $50\,\%$, which is roughly comparable to the results obtained by \citet{SmithHall11} for a slightly different definition of the WHIM and a smaller box size. However, we find around $40\,\%$ of the gas mass in the ICM of clusters, while \citet{SmithHall11} report less than $30\,\%$ mass at overdensities
greater than $10^3$ over the whole temperature range of their simulation. This is a consequence of the substantially larger box used in our simulations, which enables us to cover the high-mass range above $10^{14}\,M_\odot$ and, thus, large clusters that undergo violent mergers.\footnote{
	A similar observation can be made if the phase plot for $z=2$ in Fig~\ref{fig:phase} is compared to Fig.~5 in \citep{IapViel13}, 	which shows results from an Enzo simulation of a $10\;\mathrm{Mpc}$ box with heating, cooling, and an SGS model.} 
For halos masses $\sim 10^{13}\,M_\odot$, a substantial fraction of the gas mass can condense into the low-temperature branch, which qualifies these objects as galaxy-like. For a realistic treatment, however, cooling processes below $10^{4}\;\mathrm{K}$ and stellar feedback are necessary ingredients. This becomes obvious when volume fractions are considered (right plot in Fig.~\ref{fig:gas_frac}). The ICM and condensed warm gas fill only a very small fraction of the total volume (less than $1\,\%$). This reflects the fact that the outer shocks producing the WHIM are found at much larger radii than the high-density gas in the core (see also Section~\ref{sc:profiles}). Compared to the ICM, however, the volume of the condensed warm gas is quite large, particularly for massive clusters. This is a signature of the overly strong cooling flows produced in the high-density cores in the absence of feedback from AGNs and stars in galaxies.

\begin{table*}
  \begin{center}
      \begin{tabular}{crcccccc}
       \hline
       \multicolumn{2}{c}{} 
       & \multicolumn{3}{c}{Mass fraction} & \multicolumn{3}{c}{Volume fraction}\\
       $\sub{M}{halo}\,[M_\odot]$ & halos & 
       WHIM & ICM & cond.~warm & WHIM & ICM & cond.~warm  \\
       \hline\hline
       $>10^{14.5}$ & 8   & 0.523 & 0.381 & 0.099 & 0.9957 & 0.0022 & 0.0020\\
       $>10^{14.0}$ & 37  & 0.525 & 0.377 & 0.096 & 0.9962 & 0.0022 & 0.0016\\
       $>10^{13.5}$ & 153 & 0.518 & 0.361 & 0.106 & 0.9962 & 0.0019 & 0.0017\\
       $>10^{13.0}$ & 499 & 0.507 & 0.348 & 0.129 & 0.9961 & 0.0018 & 0.0019\\
       \hline                
      \end{tabular}
  \end{center}
  \caption{Medians of the mass and volume fractions of the ICM, WHIM, and condensed warm gas for different halo mass ranges
    (note that medians do not add up to unity). The second row indicates the total number of halos for the specified mass range.}
  \label{tb:frac}
\end{table*}

\begin{figure*}
\centering
    \includegraphics[width=\linewidth]{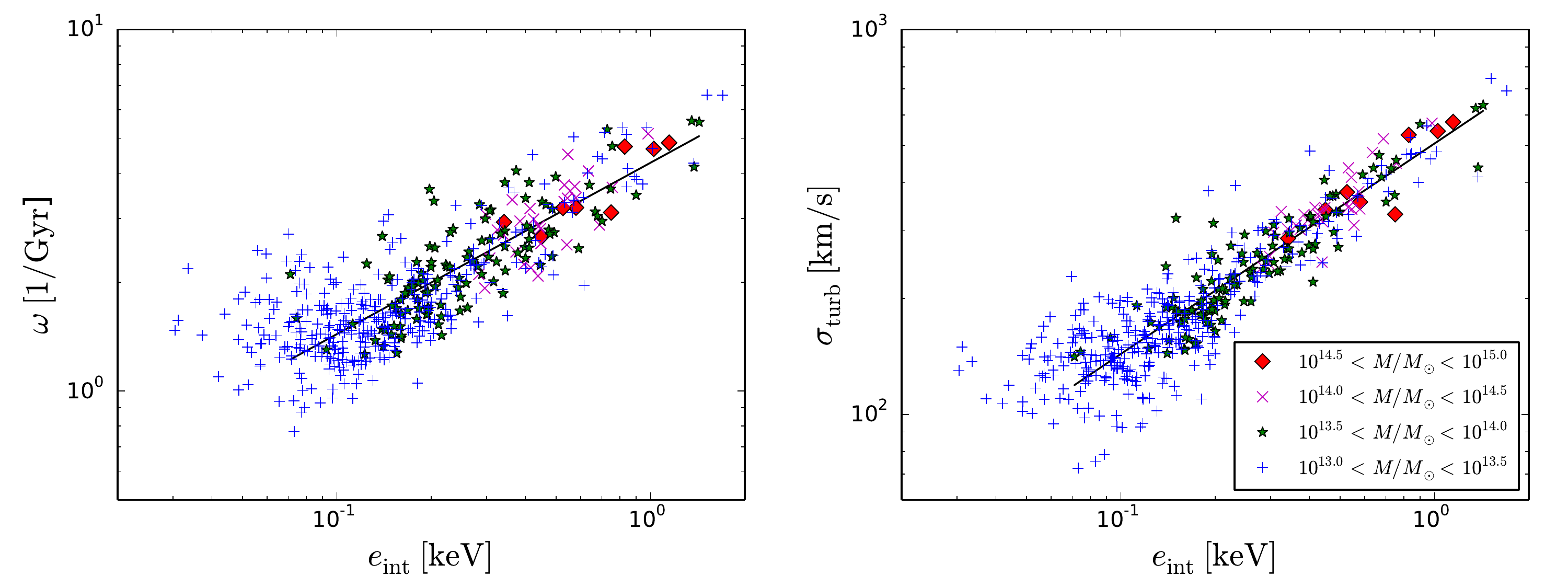}
    \caption{Mass-weighted RMS vorticity (left) and turbulent velocity dispersion (right) vs.
    	mean thermal energy for halos from run $\text{A}_1$ at redshift $z=0.0$. 
    	The different symbols correspond to the mass intervals specified in the legend. The
    	thick black lines show power-law fits to the data points with $M>10^{13.5}\,M_\odot$ 
    	(see equations~\ref{eq:mean_vort_mw} and \ref{eq:mean_sigma_mw}).}
    \label{fig:mean-mw}
\end{figure*}

\begin{figure*}
\centering
    \includegraphics[width=\linewidth]{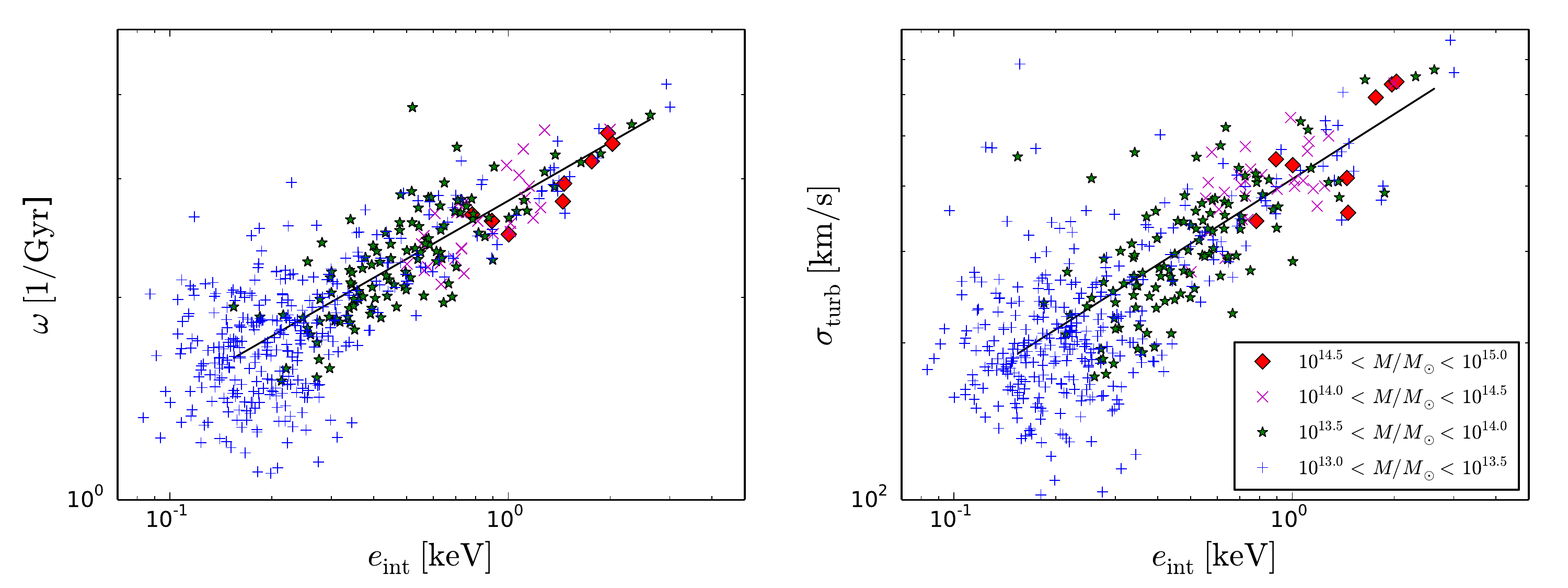}
    \caption{Volume-weighted RMS vorticity (left) and turbulent velocity dispersion (right) vs.\
    	mean thermal energy in the ICM ($\rho/\rho_0 > 500$, $T>\unit{10^5}{K}$) as in Fig.~(\ref{fig:mean-mw}).}
    \label{fig:mean-icm}
\end{figure*}

Figure~\ref{fig:mean-mw} shows the mass-weighted root-mean-square (RMS) vorticity vs.\ the mass-weighted thermal energy for a large sample of halos in simulation $\text{A}_1$. The halos are grouped in logarithmic mass intervals of $0.5\,$dex. The most massive objects with masses $M>10^{14}\,M_\odot$ correspond to clusters, the less massive objects represent groups of galaxies or giant ellipticals. Halos with masses below $10^{13}\,M_\odot$ are excluded because of the limited resolution of our simulations. There is clearly a correlation between vorticity and thermal energy, especially for the high-mass objects. For $M<10^{13.5}\,M_\odot$, however, the scatter becomes large and the slope appears to be tilted. 
A similar trend can be seen for the RMS turbulent velocity dispersion in the right plot in Figure~\ref{fig:mean-mw}. Power-law fits to the data for $M>10^{13.5}\,M_\odot$ result in the following relations:\footnote{
	For the conversion of the energy units to keV, fully ionized gas is assumed.
}
\begin{equation}
	\label{eq:mean_vort_mw}
	\langle\omega^2\rangle_\rho^{1/2} = 4.28\,\mathrm{Gyr}^{-1}\,\langle e\,[\mathrm{keV}]\rangle_\rho^{0.47}
\end{equation}
and
\begin{equation}
	\label{eq:mean_sigma_mw}
	\langle\sigma_{\rm turb}^2\rangle_\rho^{1/2} = 505\,\mathrm{km/s}\;\langle e\,[\mathrm{keV}]\rangle_\rho^{0.55},
\end{equation}	
where $\langle\ \rangle_\rho$ indicates a mass-weighted average. In both cases, the exponent is about $0.5$, which indicates that the squared turbulent velocity fluctuation is proportional to temperature. However, since gradients are dominated by fluctuations on the grid scale, the vorticity in equation~(\ref{eq:mean_vort_mw}) depends on numerical resolution. Equation~(\ref{eq:mean_sigma_mw}), on the other hand, provides a physically meaningful measure of turbulence because $\sigma_{\rm turb}$ encompasses velocity fluctuations on all scales (numerically resolved and unresolved). Consequently, the mean value of $\sigma_{\rm turb}$ is a scale-invariant integral quantity \citep[see also][]{SchmAlm14}. For some of the most massive clusters, we thus find turbulent velocity dispersions around $500\;\mathrm{km/s}$ and mean thermal energies $\sim 1\;\mathrm{keV}$. The widely spread data points for the different 
mass groups in Fig.~\ref{fig:mean-mw} also demonstrate that turbulence is only weakly correlated with the halo mass, although turbulence tends to be strong in the most massive clusters (as indicated by the red diamonds in the plots). 

\begin{figure*}
\centering
    \includegraphics[width=\linewidth]{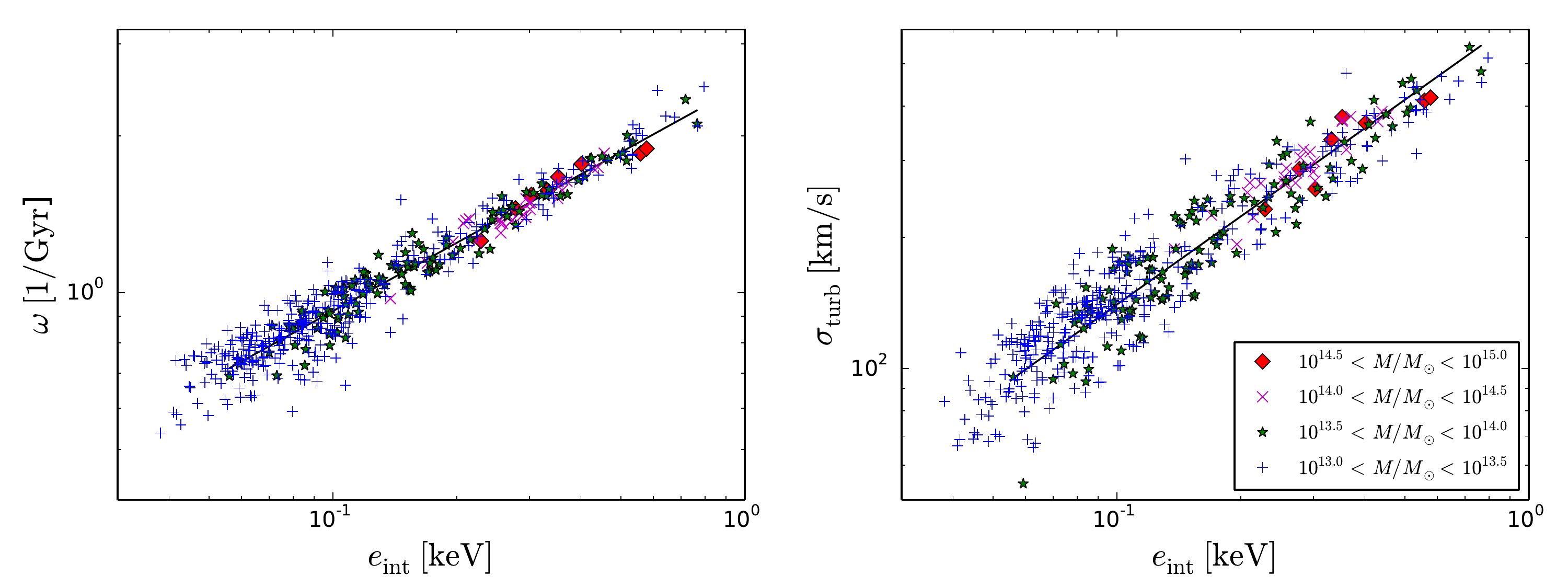}
    \caption{Volume-weighted RMS vorticity (left) and turbulent velocity dispersion (right) vs.\
    	mean thermal energy in the WHIM ($\rho/\rho_0 < 500$, $T>\unit{10^5}{K}$) as in Fig.~(\ref{fig:mean-mw}).}
    \label{fig:mean-whim}
\end{figure*}

\begin{figure*}
\centering
    \includegraphics[width=\linewidth]{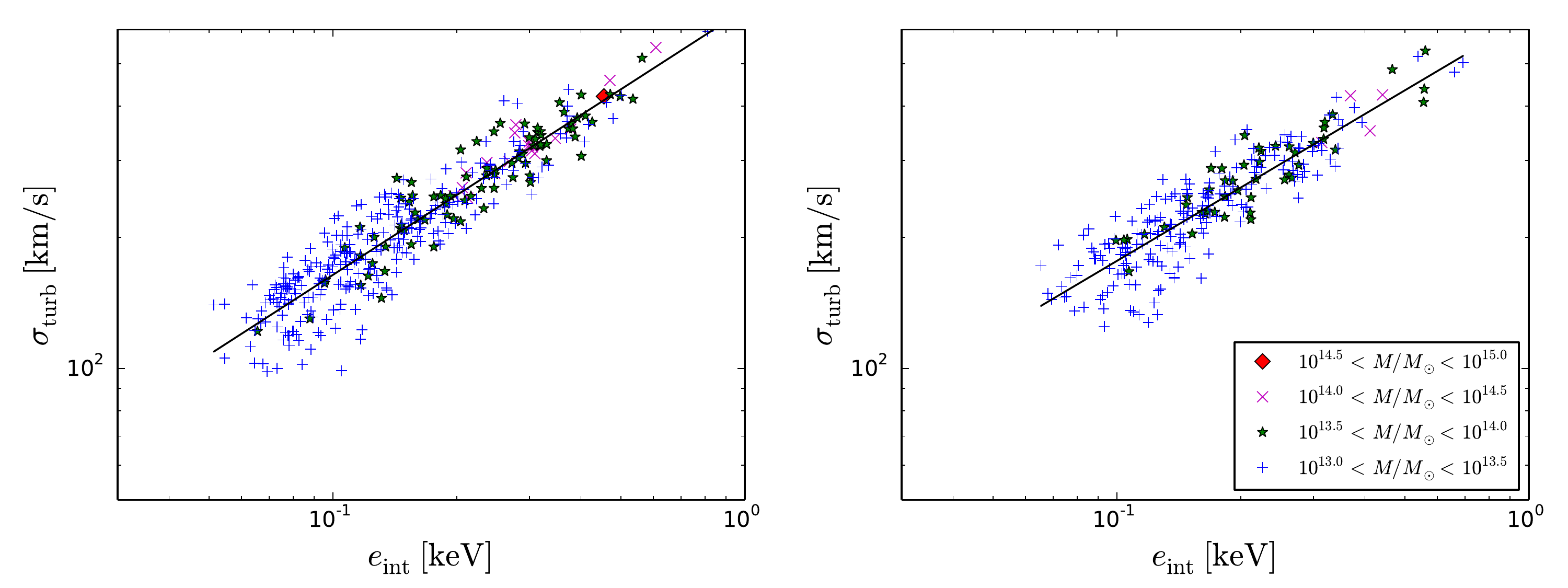}
    \caption{Volume-weighted mean turbulent velocity dispersion in the WHIM for redshifts $z=0.5$ (left)
      and $1.0$ (right).}
    \label{fig:mean-whim-sigma}
\end{figure*}

Since the gas density drops by several orders of magnitude from the 
density peak at the center of a cluster to its outskirts, a mass-weighted average samples mainly the ICM. This is confirmed by calculating volume-weighted averages with a weighing function that is constrained to the ICM, 
as defined at the beginning of Section~\ref{sc:global}. The resulting averages are plotted in Figs~\ref{fig:mean-icm} and~\ref{fig:mean-whim}. The results for the ICM are similar to the mass-weighted averages, although the scatter is larger. This is not surprising because the regions with overdensities above the chosen threshold are quite small. The fit functions for halos with $M>10^{13.5}\,M_\odot$ are 
\begin{equation}
	\label{eq:mean_vort_icm}
	\langle\omega^2\rangle_{\rm ICM}^{1/2} = 2.78\,\mathrm{Gyr}^{-1}\,\langle e\,[\mathrm{keV}]\rangle_{\rm ICM}^{0.29}
\end{equation}
and
\begin{equation}
	\label{eq:mean_sigma_icm}
	\langle\sigma_{\rm turb}^2\rangle_{\rm ICM}^{1/2} = 413\,\mathrm{km/s}\;\langle e\,[\mathrm{keV}]\rangle_{\rm ICM}^{0.41},
\end{equation}
Compared to the mass-weighted averages, we find flatter power laws for the ICM.

The volume-weighted averages for the WHIM (see Figure~\ref{fig:mean-whim}) show remarkably tight correlations, with the following fit functions:
\begin{equation}
	\label{eq:mean_vort_whim}
	\langle\omega^2\rangle_{\rm WHIM}^{1/2} = 2.51\,\mathrm{Gyr}^{-1}\,\langle e\,[\mathrm{keV}]\rangle_{\rm WHIM}^{0.44}
\end{equation}
and
\begin{equation}
	\label{eq:mean_sigma_whim}
	\langle\sigma_{\rm turb}^2\rangle_{\rm WHIM}^{1/2} = 657\,\mathrm{km/s}\;\langle e\,[\mathrm{keV}]\rangle_{\rm WHIM}^{0.67},
\end{equation}
Thus, the dependence on internal energy is stiffer than for the ICM.
In this case, more weight is put on the low-density gas, which fills a much larger volume than the ICM. Because the temperature tends to be lower in the cluster outskirts, the values of $\langle e\rangle_{\rm WHIM}$ are systematically lower than $\langle e\rangle_{\rm ICM}$. 
The WHIM contributes to mass-weighted averages, although with a lower weight than the ICM because of the smaller gas mass in the WHIM. This explains why the mass-weighted turbulent velocity dispersions are fitted by a power law (equation~\ref{eq:mean_sigma_mw}) in between the fits to the ICM and the WHIM data.

\begin{figure*}
\centering
    \includegraphics[width=\linewidth]{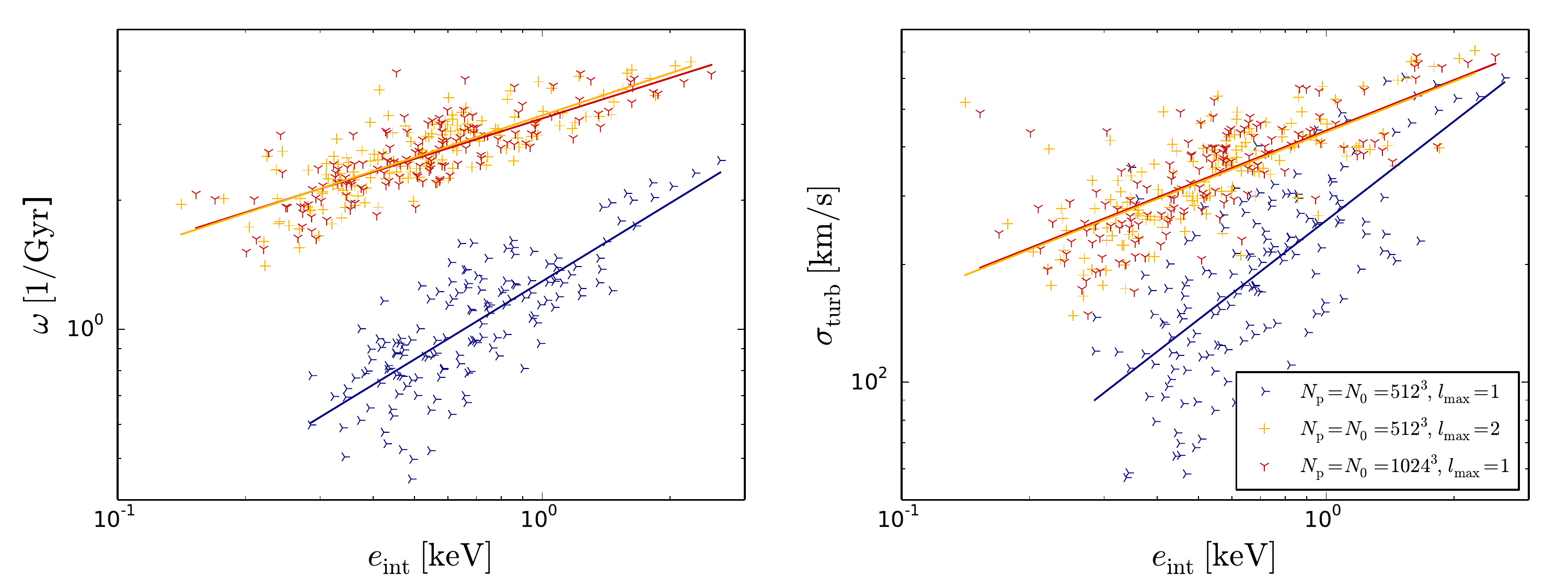}
    \caption{Volume-weighted RMS vorticity (left) and turbulent velocity dispersion (right) vs.\
    	mean thermal energy in the ICM of 153 halos in the mass range from $0.1M_{\rm max}$ to 
    	$M_{\rm max}$ for different numerical resolutions (see also Table~\ref{tb:runs}). 
    	For each data sample, a power-law fit is shown as a solid line in the same color as the data 
    	points.}
    \label{fig:mean-res-icm}
\end{figure*}

\begin{figure*}
\centering
    \includegraphics[width=\linewidth]{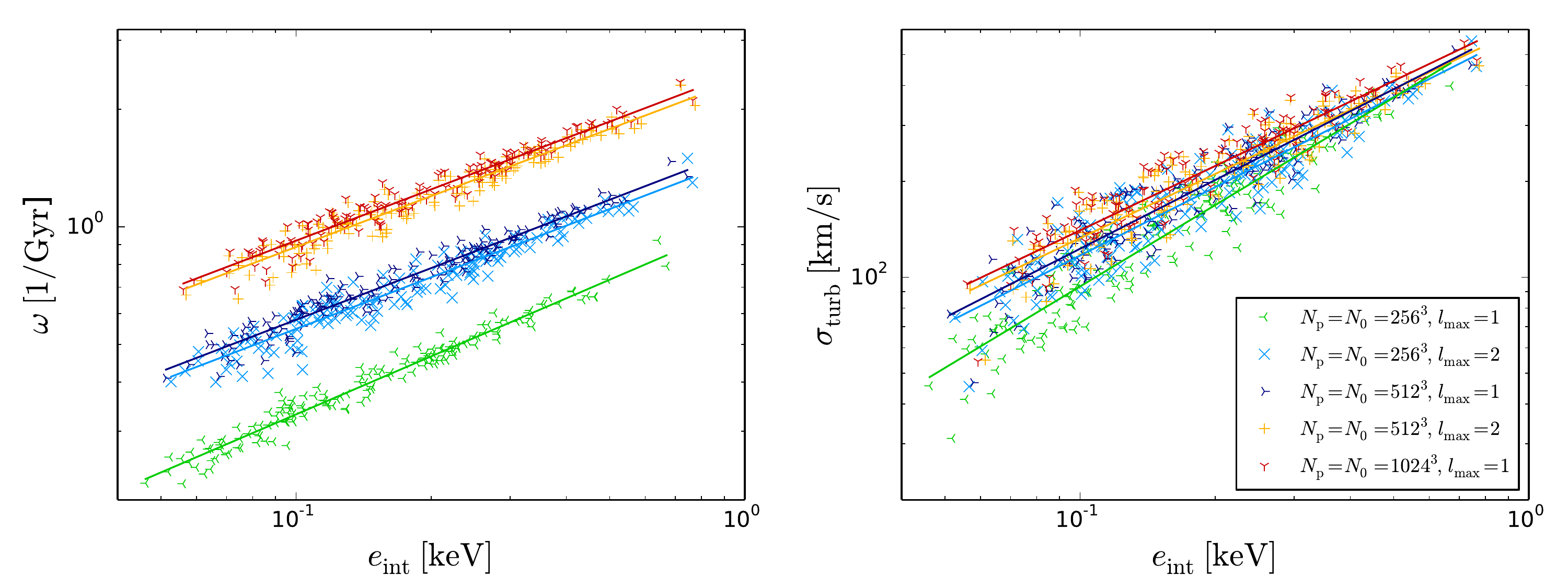}
    \caption{The same as in Fig.~\ref{fig:mean-res-icm} for the WHIM.}
    \label{fig:mean-res-whim}
\end{figure*}

At higher redshifts, we find similar power laws for the turbulent velocity dispersion in the WHIM, where $\sigma_{\rm turb}=666\,\mathrm{km/s}$ and $641\,\mathrm{km/s}$
for $e=1\,\mathrm{keV}$ at $z=0.5$ and $1.0$, respectively (see Fig.~\ref{fig:mean-whim-sigma}). The mass threshold for the fit function was lowered to $10^{13}\,M_\odot$ because of the smaller number of very massive halos at higher redshifts. 
Thus, it appears that there is no substantial evolution of the typical turbulent velocity dispersion for a given mean thermal energy. Compared to $z=0$ (right plot in Fig.~\ref{fig:mean-whim}), however, 
both the high-energy (i.~e., mean thermal energy close to $1\,\mathrm{keV}$) and low-energy tails (below $0.1\,\mathrm{keV}$) are less occupied. Particularly for halos of lower mass, the gas has not sufficiently cooled yet at $z=1$. On the other hand, extremely hot clusters also become increasingly rare at higher redshift. This agrees with the redshift-dependent distribution of the gas in the phase plots 
in Fig.~\ref{fig:phase}.

\begin{figure*}
\centering
    \includegraphics[width=\linewidth]{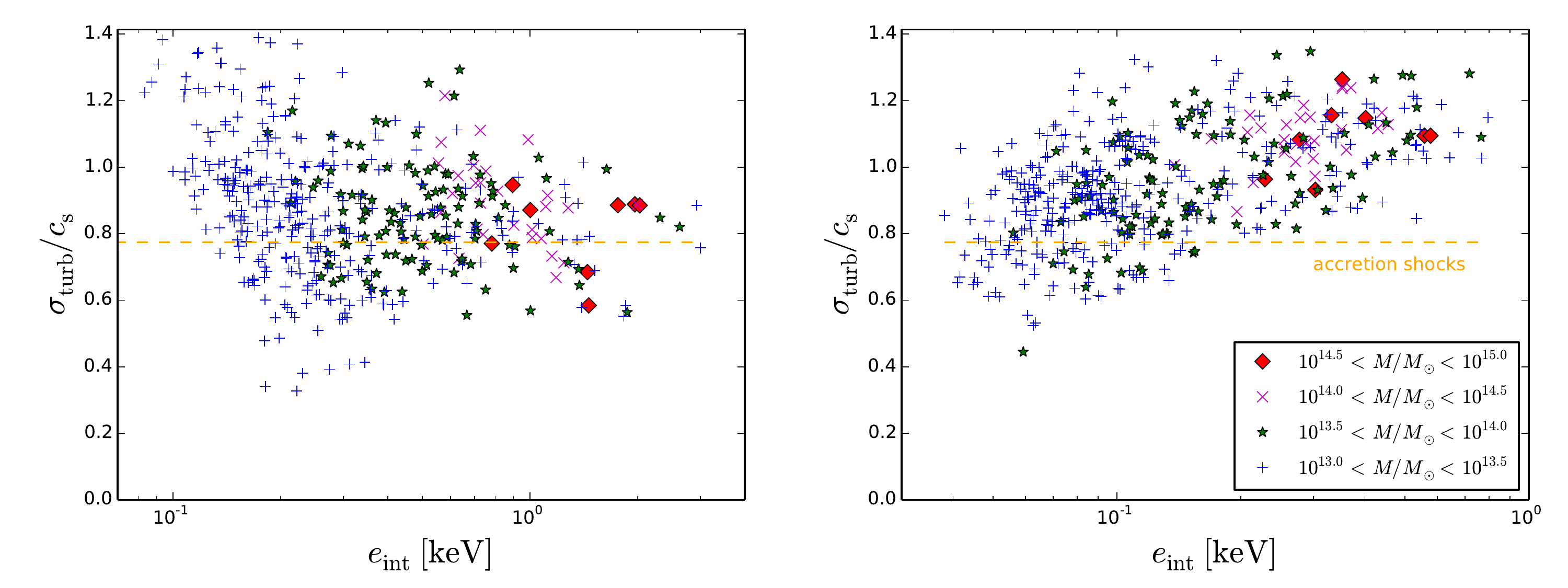}
    \caption{Volume-weighted turbulent Mach number vs. mean thermal energy for the ICM (left) and the WHIM (right) at $z=0$.}
    \label{fig:mean-mach}
\end{figure*}

\begin{table*}
  \begin{center}
      \begin{tabular}{crcccc}
       \hline
       \multicolumn{2}{c}{} & \multicolumn{2}{c}{ICM} & \multicolumn{2}{c}{WHIM}\\
        $\sub{M}{halo}\,[M_\odot]$ & halos & 
       $Q_{50\,\%}$ & $Q_{75\,\%}-Q_{25\,\%}$ &
       $Q_{50\,\%}$ & $Q_{75\,\%}-Q_{25\,\%}$ \\
       \hline\hline
       $>10^{14.5}$ & 8   & 0.88 & 0.14 & 1.09 & 0.10\\
       $>10^{14.0}$ & 37  & 0.89 & 0.16 & 1.09 & 0.10\\
       $>10^{13.5}$ & 153 & 0.87 & 0.19 & 1.01 & 0.24\\
       $>10^{13.0}$ & 499 & 0.87 & 0.24 & 0.95 & 0.23\\
       \hline                
      \end{tabular}
  \end{center}
  \caption{Medians ($Q_{50\,\%}$) and interquartiles ($Q_{75\,\%}-Q_{25\,\%}$) of the turbulent 	Mach number $\mathcal{M}_{\rm turb}$ in the ICM and WHIM for different mass ranges. }
  \label{tb:mach}
\end{table*}
 
The RMS vorticity and turbulent velocity dispersion in the ICM and WHIM are plotted
for runs with different grid and particle resolutions (see Table~\ref{tb:runs}) in Figs~\ref{fig:mean-res-icm} and \ref{fig:mean-res-whim}, respectively. Shown are data points for all halos in the mass decade below the maximal halo mass $M_{\rm max}=6.68\times10^{14}\,M_\odot$.
In the case of the ICM, both the vorticity and the turbulent velocity dispersion are
sensitive to the grid resolution scale. While this is expected for the vorticity,
the resolution dependence of the $\langle\sigma_{\rm turb}^2\rangle$ indicates that the ICM
is not properly resolved in our simulations. However, both $\omega$ and $\sigma_{\rm turb}$
suggest a trend of stronger turbulence with increasing thermal energy. Such a trend clearly follows from the data for the WHIM, for which the slopes of the power-law fits change only little with resolution. Only for the lowest spatial resolution, the fit for $\sigma_{\rm turb}$ deviates noticeably. This confirms the qualitative observation in Section~\ref{sc:runs} that the LES method breaks down for smaller halos, for which the turbulent energy is underestimated if the grid resolution is too coarse. The power-law slopes obtained for the RMS vorticity are more robust because the vorticity is directly derived from the resolved velocity field and does not rely on the Kalman filter as additional layer. However, the vorticity provides only a relative measure of turbulence, as its value is determined by the grid resolution.\footnote{
	The smaller numerical smoothing scale at higher resolution entails steeper gradients
	of the resolved turbulent velocity field if the microscopic viscosity is negligible. 
	As a result, The vorticity indefinitely increases with numerical resolution, 
	as long as the physical dissipation scale is not resolved.}
In contrast, the turbulent velocity dispersion computed with the Kalman filter approximates the magnitude of turbulent velocity fluctuations in the energy-containing range, provided that a minimum resolution is reached. The data from our simulations suggest that the required resolution is about $100\;\mathrm{kpc}$ for the WHIM in massive clusters, while higher resolution is necessary to obtain converged results for the ICM. Moreover, both quantities show a weak trend with the number of refinement levels and particle resolution $N_{\rm p}$ for a given spatial resolution $\Delta_{\rm min}$, with slightly higher values for a larger number of particles ($A_1$ vs.\ $B_2$), while no significant changes of the slopes are discernible. This might be caused by a larger, yet sub-dominant contribution from minor mergers for a higher number of dark-matter particles. The ICM data show similar trends, however, obscured by a substantially larger scatter (even in the most massive clusters, the ICM extends only over a relatively small number of grid cells in our simulations).

\begin{figure*}
\centering
    \includegraphics[width=\linewidth]{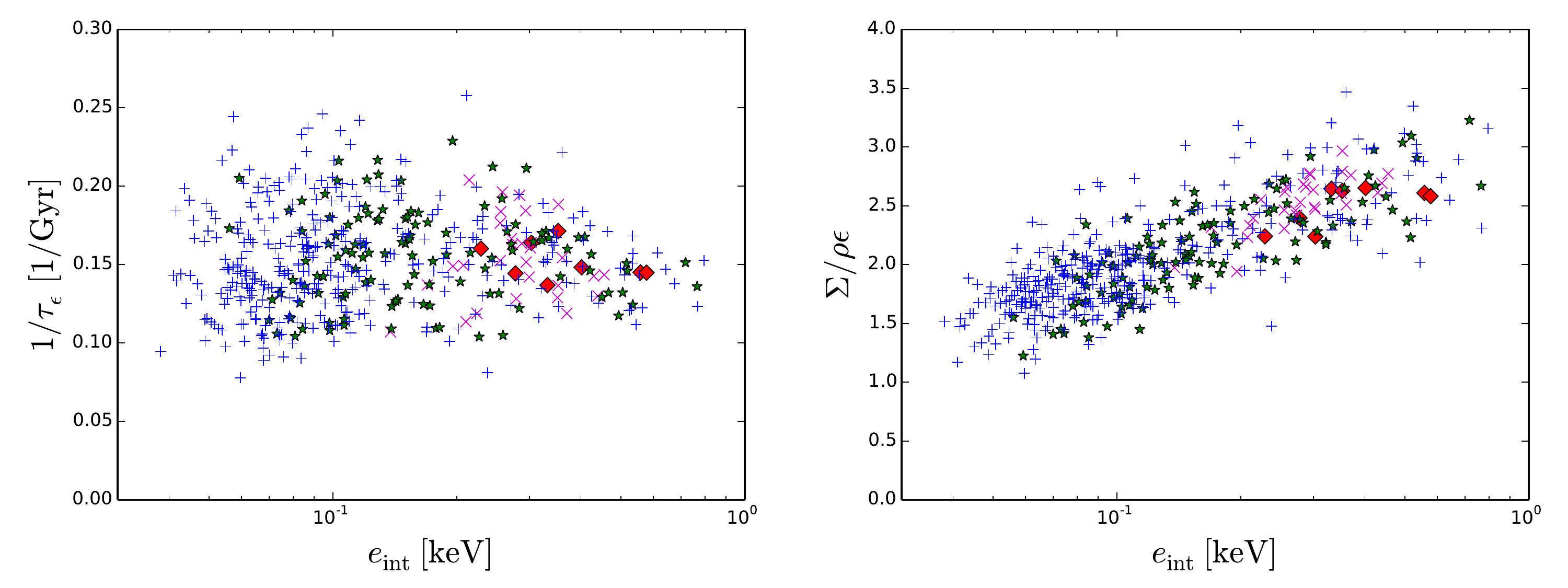}
    \caption{Volume-weighted inverse dissipation time scale (left) and production-to-dissipation ratio (right) vs. mean thermal energy for the WHIM at $z=0$.}
    \label{fig:mean-diss}
\end{figure*}

Since the relation $\sigma_{\rm turb} \propto e^{0.5} \propto c_{\rm s}$ corresponds to a constant turbulent Mach number, equations~(\ref{eq:mean_sigma_icm}) and~(\ref{eq:mean_sigma_whim}) imply that the RMS Mach number slowly decreases with the ICM temperature, but increases with the temperature of the WHIM. This is indeed demonstrated by the plots in Fig.~\ref{fig:mean-mach}, which show the turbulent Mach number defined by
\[
	\mathcal{M}_{\rm turb} = \frac{\langle\sigma_{\rm 
turb}^2\rangle^{1/2}}{\langle c_{\rm s}^2\rangle^{1/2}} =
	\left[\frac{\langle \sigma_{\rm turb}^2\rangle}{\gamma(\gamma-1)\langle e\rangle}\right]^{1/2},
\]
where $\gamma = 5/3$. For comparison, the dashed line indicates the square of the three-dimensional Mach number 
$\mathcal{M}_{\rm turb}=\sqrt{3}\mathcal{M}_{\rm s}\approx 0.77$ following from the one-dimensional jump condition for a strong adiabatic shock (equation~\ref{eq:mach1}). Since the the downstream Mach-number decreases as the adiabatic exponent approaches unity (isothermal limit, i.~e., very efficient radiative cooling), a Mach number of $0.77$ can be considered as an upper bound for shock-generated turbulence. Figure~\ref{fig:mean-mach} shows that the Mach numbers exceed this bound for many clusters, which indicates sources of turbulence production other than accretion shocks. Higher Mach numbers, however, could also result from gradual cooling of the shock-heated gas. 

\begin{table*}
  \begin{center}
      \begin{tabular}{rcccccc}
       \hline
       Rank & $\sub{M}{halo}\,[M_\odot]$ & $\sub{R}{halo}\,[\mathrm{Mpc}]$ &
       $\langle\sigma_{\rm turb}^2\rangle_{\rm ICM}^{1/2}$ & $\langle\sigma_{\rm turb}^2\rangle_{\rm WHIM}^{1/2}$ & 
       $\mathcal{M}_{\rm turb,\,ICM}$ & $\mathcal{M}_{\rm turb,\,WHIM}$ \\ 
       \hline\hline
       1 & $6.68\times 10^{14}$ & $3.24$ & 592 & 411 & 0.89 & 1.09\\
       3 & $6.18\times 10^{14}$ & $4.49$ & 627 & 366 & 0.89 & 1.15\\
       6 & $4.07\times 10^{14}$ & $5.01$ & 343 & 232 & 0.77 & 0.96\\
       10 &$2.66\times 10^{14}$ & $3.23$ & 432 & 225 & 0.99 & 1.09\\
       \hline                
      \end{tabular}
  \end{center}
  \caption{Statistics of four selected halos/clusters ($z=0.0$). From left to right: Mass rank, halo mass, maximal radial extension of the halo,
      mean turbulent velocity dispersion in the ICM and in the WHIM, turbulent Mach numbers. }
  \label{tb:halos}
\end{table*}

\begin{figure*}
\centering
    \includegraphics[width=\linewidth]{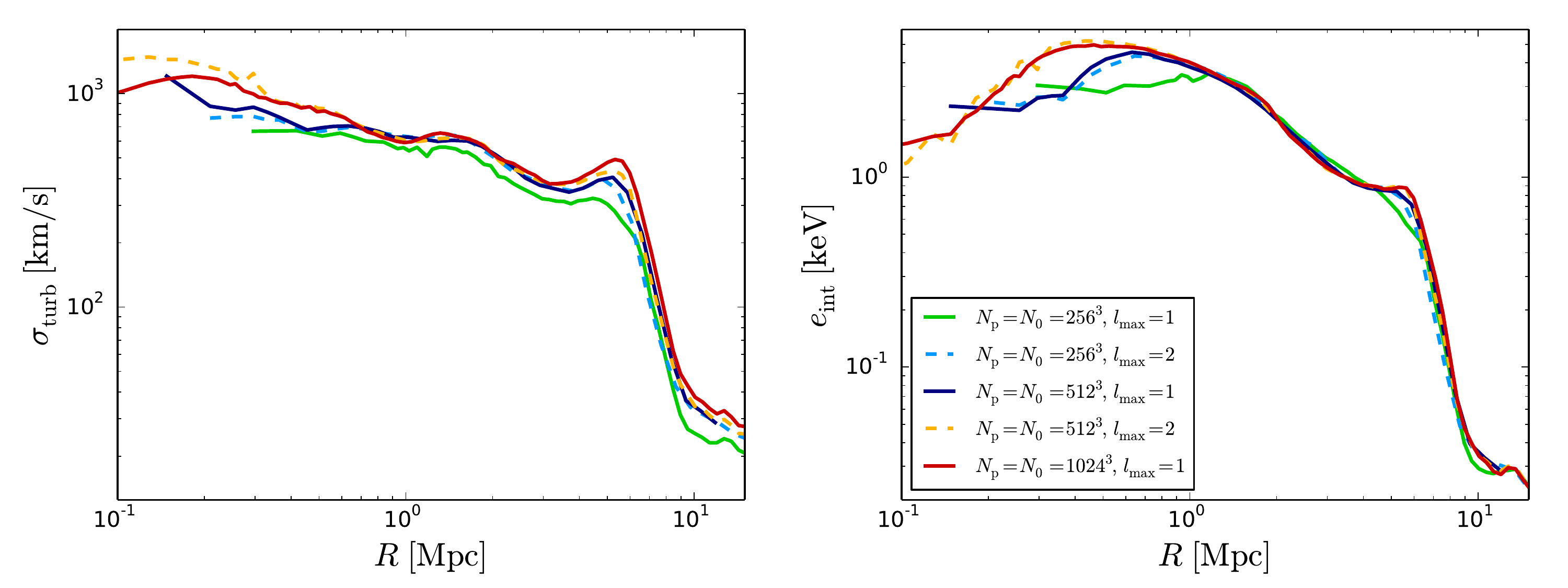}
    \caption{Radial profiles of the turbulent velocity dispersion (left) and the 	
    specific internal energy (right) for halo 3 from table~\ref{tb:halos} at different numerical resolutions.}
    \label{fig:profiles_res}
\end{figure*}

Trends also become evident by computing quartiles of the turbulent Mach numbers for samples spanning different mass ranges. The results are summarized in Table~\ref{tb:mach}. In the case of the ICM, the median (i.e., the value separating the lower and upper $50\,\%$ of data points) is remarkably constant with a value slightly below $0.9$, suggesting the relation $\sigma_{\rm turb} \propto e^{0.5}$. Although the value appears to vary with the applied method of calculation, a constant Mach number in the ICM is in agreement with the findings of \citet{VazzaBrun11} and \citet{MiniBer15}.
The the lower exponent following from the least-square fit to the ICM data (equation~\ref{eq:mean_sigma_icm}) is probably a consequence of the insufficiently resolved cluster cores (see Fig.~\ref{fig:mean-res-whim}). To better understand the origin of the relatively high Mach numbers, it is helpful to analyze the statistics for the WHIM, for which both Fig.~\ref{fig:mean-mach} and the median values reveal a trend of increasing Mach numbers for more turbulent and hotter clusters. Remarkably, the ratio of turbulence production ($\Sigma$) and dissipation ($\rho\epsilon$) plotted in Fig.~\ref{fig:mean-diss} also increases with the mean energy of the WHIM. This implies that larger clusters with higher turbulent Mach numbers experience stronger turbulence production and tend to be more out of equilibrium, i.~e., the turbulence energy flux through the cascade is not matched by dissipation rate and turbulent energy is still growing. The time scale $\tau_\epsilon=\sigma_{\rm turb}^2/(2\epsilon)$ associated turbulent energy dissipation, on the other hand, appears to be more or less constant. As a result, stronger turbulence production does not necessarily entail faster heating of the WHIM and the Mach number could indeed rise. By investigating the dependence of the effective cooling rate $\Lambda_{\rm HC}$ on the temperature at overdensities that are representative for the WHIM, it follows that $\Lambda_{\rm HC}$ increases with temperature above $10^6\;\mathrm{K}$ (roughly $0.1\;\mathrm{keV}$). If the mean temperature of the WHIM is around $10^6\;\mathrm{K}$, which is only the case for some low-mass halos in our sample, the cooling rate has a local minimum and changes only little with temperature. This implies that radiative cooling does at least not counteract the trend seen for the turbulence production rate: Stronger turbulence production and more efficient radiative cooling tend to raise the turbulent Mach number in halos with hotter WHIM. Moreover, there appears to be a plateau for halos with masses above $10^{14}\,M_\odot$, with $\mathcal{M}_{\rm turb}\approx 1.1$. This might indicate saturation, but the scatter, which is indicated by the interquartile range (i.~e., the range of values encompassed by $50\,\%$ of the data points centered around the median), is too large to ascertain such a subtle change in the trend.

\section{Cluster profiles}
\label{sc:profiles}

To investigate the properties of individual clusters, we computed radial profiles. Table~\ref{tb:halos} lists a sample of four clusters, which are are among the top ten in mass. To begin with, radial profiles are probes of the sensitivity on resolution.  As representative example, profiles of the two key quantities of our analysis, the turbulent velocity dispersion $\sub{\sigma}{turb}$ and the thermal energy $e$, are plotted for a particular halo in Fig.~\ref{fig:profiles_res}. The profiles of $\sub{\sigma}{turb}$ agree very well for the different runs, which corroborates the results shown in Fig.~\ref{fig:mean-res-whim}. While the thermal energy of the cluster outskirts are also very robust, we find a strong sensitivity to the grid resolution for the ICM at radii smaller than about $1\;\mathrm{Mpc}$. Basically, cooling creates a central depression in the temperature profile, which becomes deeper and narrower as the grid resolution increases. As a consequence, significantly higher core resolution would be required. Moreover, only the incorporation of feedback processes can ensure that overcooling is prevented and converged temperature profiles are obtained in the cluster core \citep{BorKravt11,KravtBor12,HahnMar15}.
However, it is important to notice that the volume, for which the thermal energy is nearly independent of resolution, is of the order $10^3$ times the innermost Mpc$^3$.

\begin{figure*}
\centering
    \includegraphics[width=\linewidth]{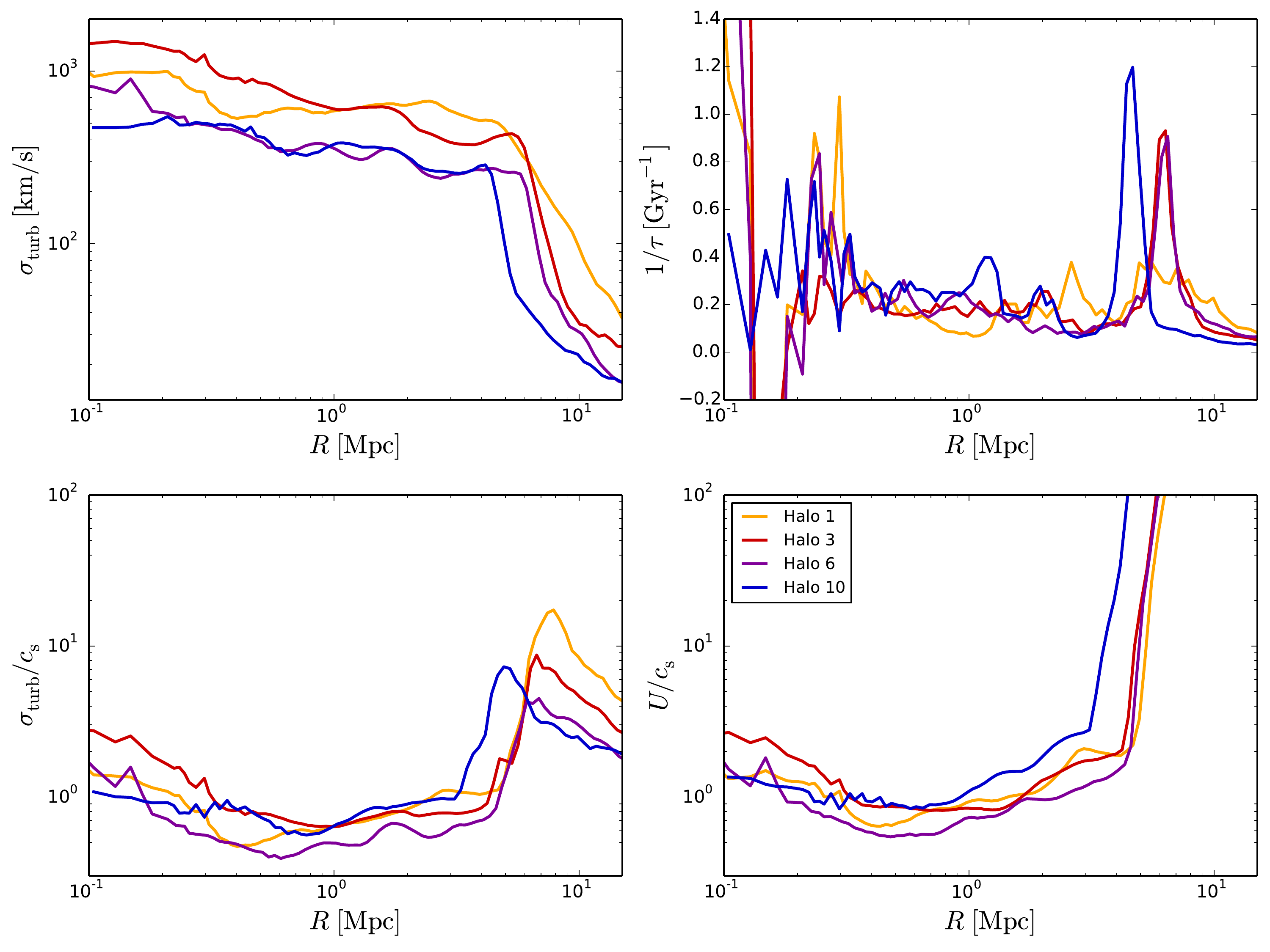}
    \caption{Radial profiles of the turbulent velocity dispersion (top left), the inverse dynamical
    	time scale (top right), the turbulent Mach number (bottom left), and the total Mach number (bottom right)
	for four selected halos at $z=0$ (run B2). The halos are ranked by mass in the legend (see Table~\ref{tb:halos}).}
    \label{fig:profiles_turb_z0}
\end{figure*}

For $\sub{\sigma}{turb}$, we find the typical shape that was reported by \citet{SchmAlm14} for the Santa Barbara cluster: The profile of $\sub{\sigma}{turb}$ is rather flat in the interior of each cluster and rapidly drops at the accretion shocks, which are located at radii of almost \unit{10}{Mpc}  (the shocked gas extends further outwards than the radius $\sub{R}{halo}$ of the dark matter halo). The turbulent velocity dispersion of these large clusters can reach \unit{1000}{km/s} close to the center, whereas the mean values listed in Table~\ref{tb:halos} are not quite as large. The dynamical time scale $\tau$  defined by equation~(\ref{eq:tau_dyn}) assumes values around \unit{5}{Gyr} for $R\sim\unit{1}{Mpc}$ for all clusters (top right plot in Fig.~\ref{fig:profiles_turb_z0}). The pronounced peaks of $1/\tau$ between $4$ and $\unit{8}{Mpc}$ are induced by accretion shocks. This demonstrates that turbulence is indeed efficiently generated by shocks, as previously shown by \citet{KangRyu07,RyuKang08,VazzaBrun09b,PaulIap11,Miniati2014}. Since \citet{KangRyu07} and \citet{RyuKang08} define a dynamical time scale $\tau$ in terms of the vorticity,
their statement that $\tau$ assumes a typical value in terms of the Hubble time depends on the chosen grid scale. In contrast, equation~(\ref{eq:tau_dyn}) provides a definition independent of the grid resolution \citep[see][]{SchmAlm14}. In the center of the clusters, the production rate $\Sigma$ (see equation~\ref{eq:prod}) can become negative, which implies strong backscattering from subgrid to resolved scales. The strong fluctuations at small radii are a consequence of the small number of grid cells, over which data are averaged. The bottom plots show profiles of the Mach numbers $\sub{\sigma}{turb}/\sub{c}{s}$ and $U/\sub{c}{s}$ associated with turbulence and the resolved flow, respectively. While the turbulent Mach numbers are mostly subsonic or transonic in the clusters, there is a steep rise to supersonic values around the accretion shocks, which is overcompensated by the rapidly decreasing turbulent velocity fluctuation at larger radii. The Mach numbers associated with the total velocity $U$, on the other hand, increase to extremely high values \citep[see][]{BorKravt11}. This is caused by the contribution from the accretion of cold gas, which falls with very high speed toward the massive clusters. 

\begin{figure*}
\centering
    \includegraphics[width=\linewidth]{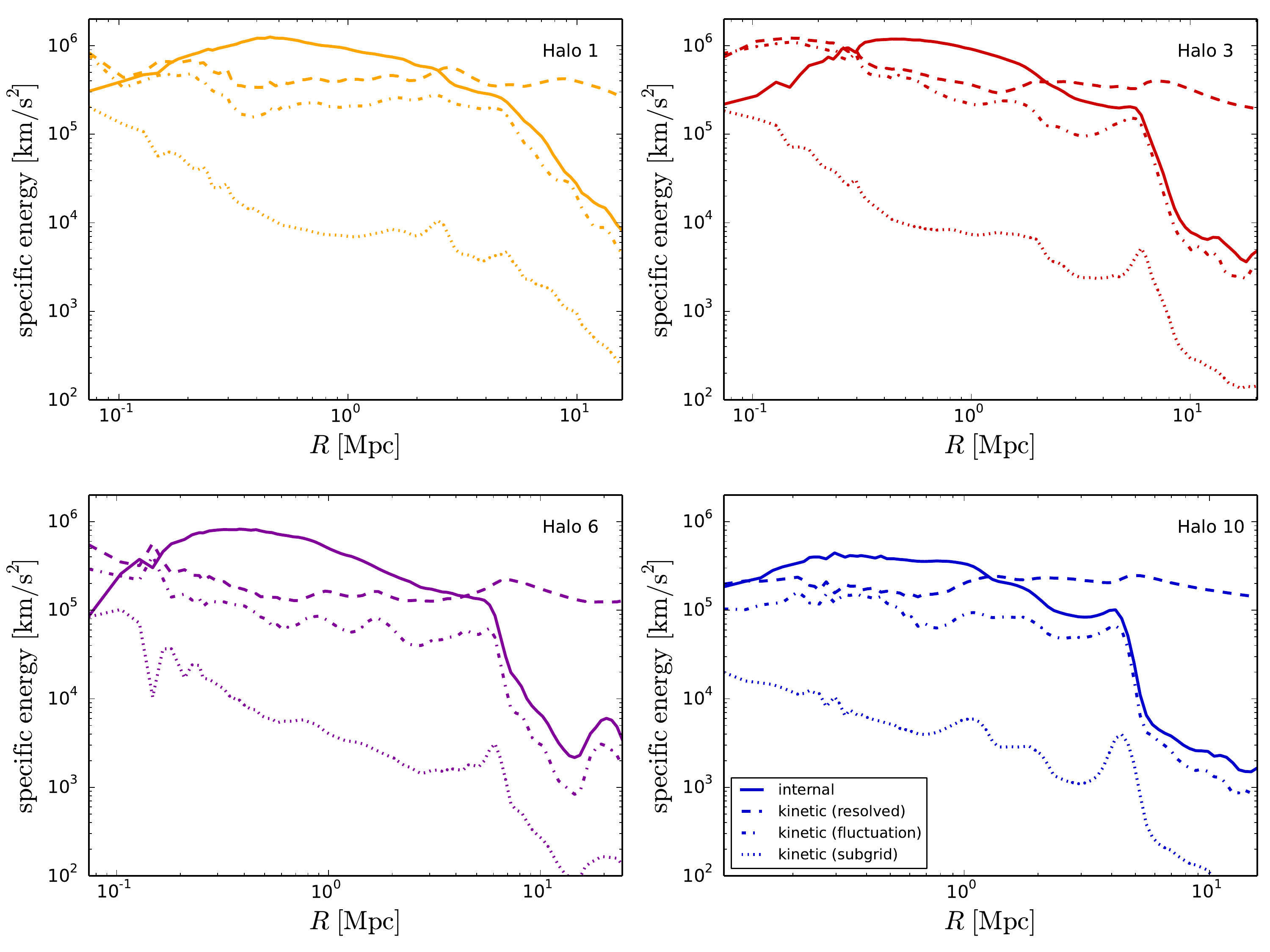}
    \caption{Radial profiles of the internal and kinetic energies per unit mass for the halos listed in Table~\ref{tb:halos}.
      For the kinetic energy, the numerically resolved contribution, $U^2/2$ (dashed lines), the energy associated with the turbulent
      velocity fluctuation, $U^{\prime\,2}/2$ (dot-dahsed lines), and the subgrid-scale contribution K (dotted lines) are plotted.}
    \label{fig:profiles_energy_z0}
\end{figure*}

The different contributions to the specific energy are plotted for the four selected halos in Fig.~\ref{fig:profiles_energy_z0}.
By comparing $U^2/2$ (total resolved kinetic energy) and $U^{\prime\,2}/2$ (resolved turbulent energy), it can be seen that a large fraction of the kinetic energy is turbulent inside the accretions shocks, but dominated by bulk motions in the exterior medium. The SGS energy $K$ contributes a small, but non-negligible fraction of the turbulent kinetic energy. Moreover, $U^{\prime\,2}\sim e$ throughout the clusters, which is in agreement with mean Mach numbers of the order unity in the WHIM (see Table~\ref{tb:halos}).
However, efficient cooling in the cluster core substantially reduces the temperature \citep{LukStark15}, producing a characteristic dip in the internal energy profile, which is particularly pronounced for halos 3 and 6 in Fig.~\ref{fig:profiles_energy_z0}, and an increase of the Mach number toward the center (see bottom left plot in Fig.~\ref{fig:profiles_turb_z0}). This also affects the mean Mach numbers of the ICM. The flatter internal energy profiles for or halos 1 and 10 reflect the heat released by recent major mergers.

\begin{figure*}
\centering
    \includegraphics[width=\linewidth]{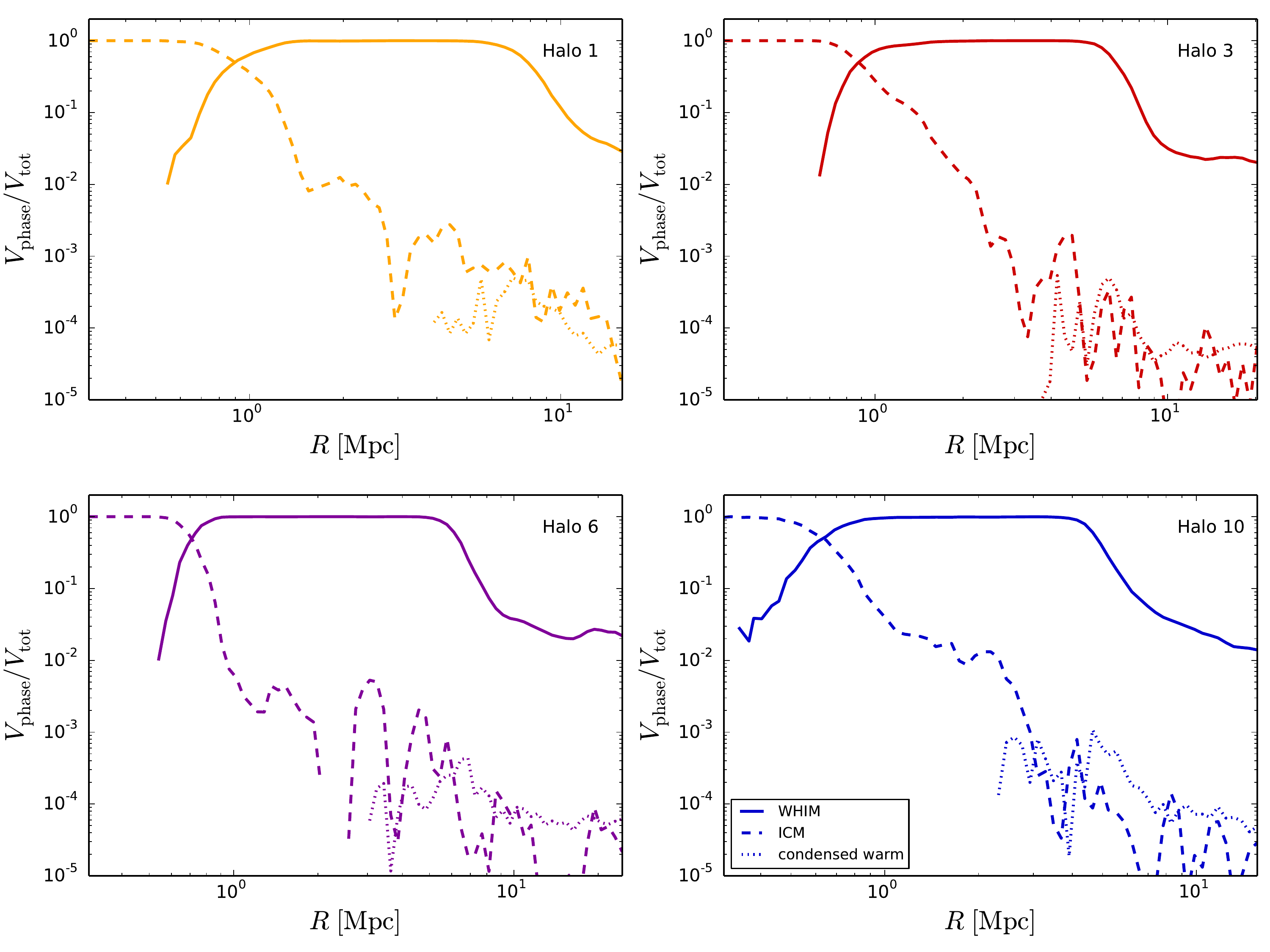}
    \caption{Radial profiles of the volume fractions $V_{\rm phase}/V_{\rm tot}$ of the 
    WHIM, ICM, and condensed warm gas for the same halos as in Fig.~\ref{fig:profiles_energy_z0},
    where $V_{\rm tot}$ is the total volume of a radial shell.}
    \label{fig:profiles_fract_z0}
\end{figure*}

Figure~\ref{fig:profiles_fract_z0} shows how the distribution of the gas among the phases defined in Sect.~\ref{sc:global} changes with radius. In contrast to the mean gas fractions plotted in Fig.~\ref{fig:gas_frac}, the volume fractions are normalized to the total volume, including low-density gas in the void. The WHIM fills nearly all the cluster volume enclosed by the outer shocks. The residual WHIM fraction outside of the accretion shocks stems from other nearby clusters or filaments extending further outwards. At radii between $0.6$ and \unit{0.9}{Mpc}, the ICM fraction crosses the WHIM fraction. Although there is no sharp separation between WHIM and ICM, the cluster core is clearly dominated by the ICM. The applied definitions are therefore sensible. Halo 10 is unusual, as a non-negligible amount of warm-hot gas is mixed into the core, which is another indication of recent merging. Moreover, for all halos the ICM fraction is does not monotonically decrease with radius. There are peaks corresponding to small volume fractions of an ICM-like phase and condensed warm gas at radii of several Mpc, which can be attributed to subhalos merging with the cluster (inside the accretion shock) or subhalos in the vicinity of the cluster.

\section{Conclusions}
\label{sc:concl}

We analyzed thermal and turbulent properties of the baryonic gas in galaxy clusters by running simulations of cosmic structure formation with the method of adaptively refined large eddy simulations introduced in \citet{SchmAlm14}. We used the treatment of radiative heating and cooling of \citet{LukStark15} to evolve baryonic gas from initial conditions produced with \textsc{Music} \citep{HahnAbel11}, assuming the $\Lambda$CDM parameters established by the Planck mission. 
We neglected feedback processes, which foremost rely on heuristics. This allowed us to apply relatively well known and numerically controllable physics. As an indicator of turbulence, the fluctuating component of the numerically resolved peculiar velocity is calculated with a Kalman filtering algorithm. In addition, the numerically unresolved component is determined by a subgrid scale model for compressible turbulence \citep{SchmFeder11}. Both components are used to obtain an estimate of the turbulent velocity dispersion $\sigma_{\rm turb}$ in clusters. As opposed to previous methods for approximating the turbulent velocity and associated time scales in clusters \citep[e.~g.][]{NorBry99,DolVaz05,KangRyu07,MaierIap09,ZhuFeng10}, our definition of $\sigma_{\rm turb}$ respects the three-dimensional nature of turbulence, does not depend on spherical symmetry, nor is it subject to a particular choice of a spatial smoothing length or resolution scale.

The radial profiles of $\sigma_{\rm turb}$ computed with the Kalman filter and shear-improved subgrid-scale model are rather flat within a radius of a few Mpc from the density peak of a cluster, with a sharp drop at the outer shocks (see Figs~\ref{fig:profiles_turb_z0}).
For relaxed clusters, we also see a strong peak of the inverse dynamical time scale associated with the turbulent cascade (basically, the turbulence energy flux normalized by the energy). This indicates that accretion shocks not only heat the infalling gas to high temperatures but also generate turbulence. Indeed, radial profiles of the specific energy show that both the thermal and turbulent energies are substantially raised (Fig.~\ref{fig:profiles_energy_z0}). Qualitatively, our findings are in agreement with previous studies by \citet{RyuKang08,LauKravt09,PaulIap11,VazzaBrun11,Miniati2014} and others.

\begin{figure}
\centering
    \includegraphics[width=\linewidth]{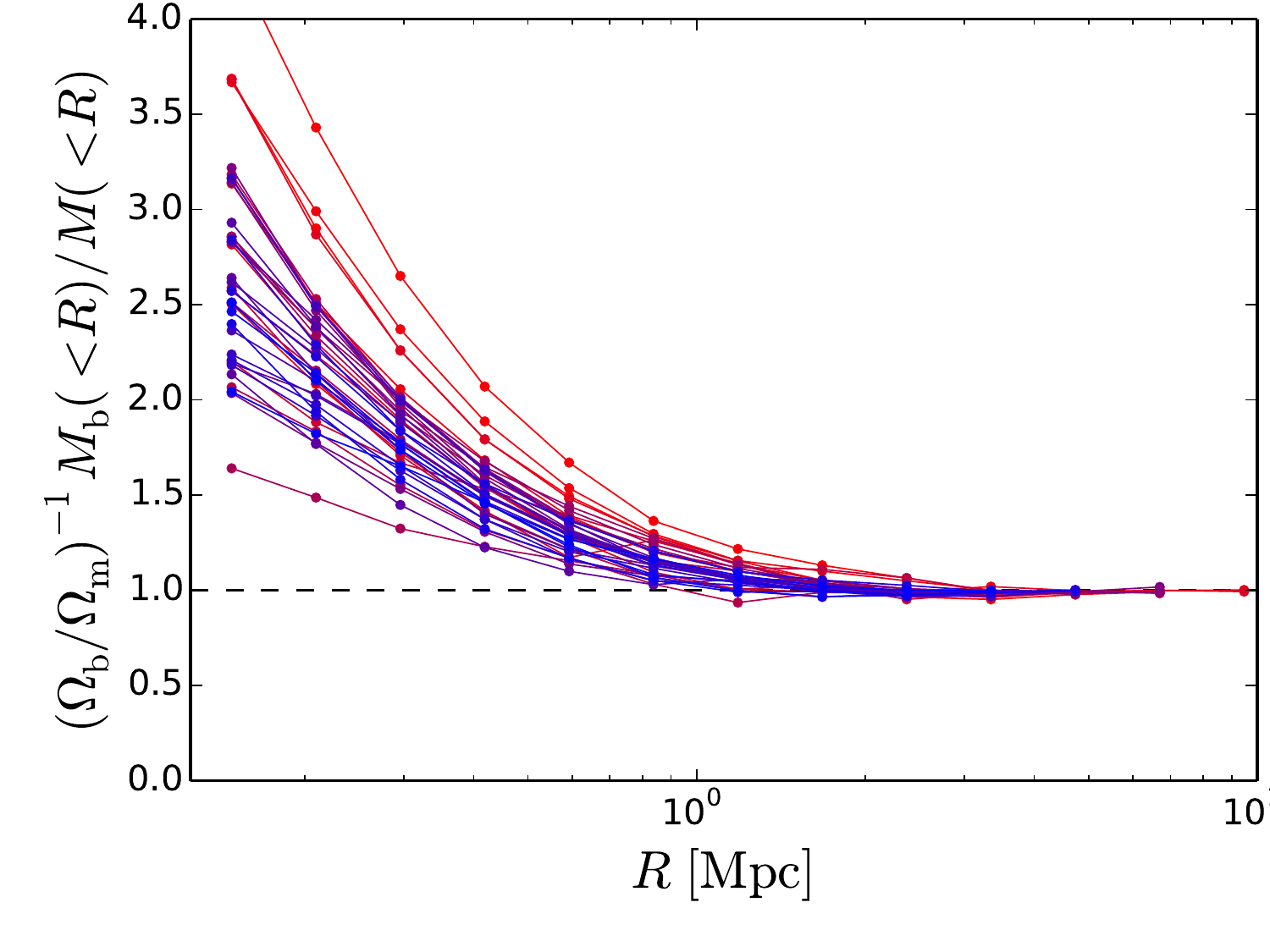}
    \caption{Cumulative baryon fractions for the 37 clusters with 
    $\sub{M}{halo}>10^{14}\,M_\odot$ (most massive in red) relative to the cosmological baryon fraction 
    $\Omega_{\rm b}/\Omega_{\rm m}=0.155$ at $z=0$. $M(<\!R)$ is the total mass within
    a sphere of radius R and $\sub{M}{b}(<\!R)$ the corresponding mass of baryonic matter.}
    \label{fig:baryon_fraction}
\end{figure}

A global phase diagram of the turbulent Mach number associated with $\sigma_{\rm turb}$ vs.\ overdensity shows a supersonic and a transonic branch at high density (see Fig.~\ref{fig:phase_mach}). The former is absent if only adiabatic gas dynamics is applied \citep[see Fig.~20 in][]{SchmAlm14} and can be understood as a consequence of radiative cooling. Moreover, the Mach numbers in the low-density gas tend to be lower because of the UV background \citep{VazzaBrun09a}. The effect of cooling is also reflected by a central dip of the thermal energy in cluster profiles. Of course, feedback from galaxies affects both the thermal and turbulent energies of IGM. Although AGN feedback and supernova-driven winds cease to be major sources of energy at low redshift, the cumulative mass fractions of baryons plotted in Fig.~\ref{fig:baryon_fraction} are indicative of strong overcooling in the cores of all massive clusters in our simulations, i.e. the pressure support of the gas against gravity is too weak. \citet{HahnMar15} demonstrated that AGN feedback is essential to reproduce observed properties of the ICM in cluster cores. However, it is not completely clear yet how a consistent treatment of turbulent energy injection, independent of the grid resolution, can be achieved in cosmological simulations \citep[see, for example,][]{BrueggScann09,BorKravt11,YangSut12,VazzaBruegg13,HahnMar15}.

To analyze the turbulent velocity dispersion and Mach numbers in different clusters, we applied a halo finder and computed mean values for the gas associated with the dark matter halos. The gas is separated into the ICM defined by baryonic overdensities $\rho/\rho_0>500$ and the WHIM with $\rho/\rho_0<500$. In both cases a temperature threshold of $T=10^5\,\mathrm{K}$ is applied. We find power laws for the mean turbulent velocity dispersion as a function of the mean thermal energy of the gas (equations~\ref{eq:mean_sigma_icm} and~\ref{eq:mean_sigma_whim}), where the slope is about $0.4$ for the ICM, while the relation for the WHIM is stiffer, with an exponent of almost $0.7$ (the slope would be $0.5$ if the Mach numbers were constant for all clusters). A power law with an intermediate slope follows for the mass-weighted mean values. 
However, the slope for the ICM is not converged in our simulations because of the lack of resolution in the cluster cores, which is severely limited if refinement on the magnitude of the vorticity is applied throughout the domain. The resolution trend suggests that the correct exponent is close to $0.5$, indicating a constant ratio of thermal and turbulent energies in the ICM.
At redshift $z=0$, the highest mean thermal energies of the ICM are of the order $1\,\mathrm{keV}$ and the corresponding turbulent velocity dispersion is roughly $400\,\mathrm{km/s}$. Values above $500\,\mathrm{km/s}$ are found in rare cases. 
Such clusters are typically very massive, but the scatter of the turbulent velocity dispersions in terms of the dark halo mass is very large, particularly at the lower-mass end. The turbulent velocity dispersions $\sigma_{\rm turb}^2\sim 10^5\ldots 10^6\,\mathrm{(km/s)}^2$ in our simulations are comparable to the integral-scale velocity fluctuations following from the computation of second-order structure functions by \citet{Miniati2014,Miniati2015}. \citet{VazzaBrun11,VazzaBruegg13}, on the other hand, calculated turbulent velocities that are lower by factors of about $2$ to $3$. This might be a consequence of the preferred scales of their filtering method, which tend to be smaller than the integral scales reported by \citet{Miniati2015}. Their method and ours, however, are likely to bracket the physical turbulent velocity dispersion that is characteristic for clusters. 

The mean turbulent Mach numbers in the ICM lie mainly in the subsonic range, particularly for hot and massive clusters (see Fig.~\ref{fig:mean-mach}). The expected Mach number would be $0.8$ or less if turbulence were solely generated by accretion shocks. However, there are also halos with higher Mach numbers, which can result from cooling in denser regions and turbulence driven by mergers.
The turbulent Mach numbers implied by \citet{VazzaBrun11} are below $0.7$. Similar or even lower values were obtained, for example, by \citet{LauKravt09,Valdar11,PaulIap11}. Apart from the different methods of calculating the turbulent velocity dispersion, the deviations might be caused by systematics related to the simulations, such as the lack of radiative cooling or different box sizes favoring less violent evolution. On the other hand, our results agree very well with the turbulent Mach numbers calculated by \citet{MiniBer15}, which suggests that structure functions yield similar results for the turbulent velocity dispersion as our approach. In the WHIM, we typically find transonic or mildly supersonic turbulence. There is a weak trend of higher Mach numbers in hotter gas, which requires an overcompensation of the higher thermal energy by turbulence energy injection.  This is indeed seen for the mean contribution of turbulence production relative to dissipation into thermal energy.

At least for the ICM, an observational determination of the relation between thermal and turbulent energies should become feasible with high-precision X-ray spectroscopy \citep{SandFab11,ZhurChur11}. However, relations such as equation~(\ref{eq:mean_sigma_mw}) must be interpreted with care in an observational context.
First of all, the step from the mass-weighted or constrained averages applied in our work to the temperature and density dependence of X-ray emissions is far from straightforward. Another issue is that observed line broadening in clusters typically selects a scale that is given by the angular resolution of the instrument, while we consider quantities that are by definition scale invariant (with some residual dependence on numerical resolution). A link could be provided by the subgrid scale energy, which specifies turbulent velocity fluctuations below a particular scale. This scale is given the local grid resolution. In AMR simulations, subgrid scale energies are available for the whole grid hierarchy. Thus, one could interpolate to any given scale bracketed by the resolution of the highest refinement level and the root grid. 
If cosmological simulations are performed with many refinement levels, the accessible range of length scales will be $\sim 10\,\mathrm{kpc}$ to $\sim 1\,\mathrm{Mpc}$. This encompasses typical spatial resolutions of upcoming cluster observations, for example, with Astro-H.

\section*{Acknowledgments}

We thank Peter Nugent and his team for supporting the development of \nyx\ at the Computational Cosmology Center at LBNL.
In particular, we thank Zarija Luki\'{c} for the implementation of heating and cooling.
Moreover, we thank Marcus Br\"uggen for useful comments, particularly with regard to observational aspects. 
W.~Schmidt, J.~F.~Engels, and J.~C.~Niemeyer acknowledge financial support by the CRC 963 of the German Research Council. 
The work at LBNL was supported by the SciDAC Program and the Applied Mathematics Program of the U.S. Department of Energy under Contract No. DE-AC02-05CH11231.
The simulations presented in this article were performed on the HLRN-III complex \emph{Gottfried} in Hannover (project nip00034). 
We also acknowledge the yt toolkit by \citet{TurkSmith11} that was used for our analysis of numerical data. 

\bibliography{HotTurbClusters}

\label{lastpage}

\end{document}